\documentclass{article}

\usepackage{PRIMEarxiv}

\usepackage[utf8]{inputenc} % allow utf-8 input
\usepackage[T1]{fontenc}    % use 8-bit T1 fonts
\usepackage{hyperref}       % hyperlinks
\usepackage{url}            % simple URL typesetting
\usepackage{booktabs}       % professional-quality tables
\usepackage{amsfonts}       % blackboard math symbols
\usepackage{nicefrac}       % compact symbols for 1/2, etc.
\usepackage{microtype}      % microtypography
\usepackage{lipsum}
\usepackage{fancyhdr}       % header
\usepackage{graphicx}       % graphics
\graphicspath{{media/}}     % organize your images and other figures under media/ folder

\usepackage{times}  % DO NOT CHANGE THIS
\usepackage{helvet}  % DO NOT CHANGE THIS
\usepackage{courier}  % DO NOT CHANGE THIS
\usepackage{graphicx} % DO NOT CHANGE THIS
\usepackage{caption} % DO NOT CHANGE THIS AND DO NOT ADD ANY OPTIONS TO IT
\usepackage{times}
\usepackage{soul}
\usepackage{url}
\usepackage[utf8]{inputenc}

\usepackage{graphicx}
\usepackage{amsmath}
\usepackage{amsthm}
\newtheorem{definition}{Definition}
\usepackage{amsfonts}
\usepackage{mathrsfs}
\usepackage{makecell}
\usepackage{booktabs}
\usepackage{algorithm}
\usepackage{algpseudocode}
\usepackage{multirow}
\usepackage[switch]{lineno}
\usepackage{newfloat}
\usepackage{listings}

\usepackage{nicefrac}       % compact symbols for 1/2, etc.
\usepackage{microtype}      % microtypography
\usepackage{lipsum}
\usepackage{fancyhdr}       % header                   
\usepackage{subcaption}   
\usepackage{cleveref}
\usepackage{wrapfig}
\usepackage{siunitx}
\usepackage{bbm}

\usepackage[x11names]{xcolor}

\newtheorem{theorem}{Theorem}

\title{Leveraging Side Information for Ligand Conformation Generation using Diffusion-Based Approaches

}

\author{
  Jiamin WU, He CAO \\
  Department of Mathematics\\
  Hong Kong University of Science and Technology \\
  Hong Kong\\
  \texttt{\{jwubz, hcaoaf\}@connect.ust.hk} \\
   \And
  Yuan YAO \\
  Department of Mathematics\\
  Hong Kong University of Science and Technology \\
  Hong Kong\\
  \texttt{\{yuany\}@ust.hk}
}

\begin{document}
\maketitle

\begin{abstract}
Ligand molecule conformation generation is a critical challenge in drug discovery. Deep learning models have been developed to tackle this problem, particularly through the use of generative models in recent years. However, these models often generate conformations that lack meaningful structure and randomness due to the absence of essential side information. Examples of such side information include the chemical and geometric features of the target protein, ligand-target compound interactions, and ligand chemical properties. Without these constraints, the generated conformations may not be suitable for further selection and design of new drugs. To address this limitation, we propose SIDEGEN, a novel method for generating ligand conformations that leverage side information and incorporate flexible constraints into standard diffusion models. SIDEGEN employs center of mass and equivariant transformation techniques, which ensure translational and rotational invariance in Euclidean space. Drawing inspiration from the concept of message passing, we introduce ligand-target massage passing block (LTMP), a mechanism that facilitates the exchange of information between target nodes and ligand nodes, thereby incorporating target node features. To capture non-covalent interactions, we introduce ligand-target compound inter and intra edges. To further improve the biological relevance of the generated conformations, we train energy models using scalar chemical features, including Self-consistent field energy, molecular orbital–lowest unoccupied molecular orbital energy gaps, and Marsili-Gasteiger Partial Charges. These models guide the progress of the standard Denoising Diffusion Probabilistic Models, resulting in more biologically meaningful conformations. We evaluate the performance of SIDEGEN using the PDBBind-2020 dataset, comparing it against other methods. The results demonstrate improvements in both Aligned RMSD and Ligand RMSD evaluations. Specifically, SIDEGEN outperforms GeoDiff (trained on PDBBind-2020) by 20\% in terms of the median aligned RMSD metric.

\end{abstract}

\begin{figure}[!hb]
  \centering
  \includegraphics[width=0.95\columnwidth]{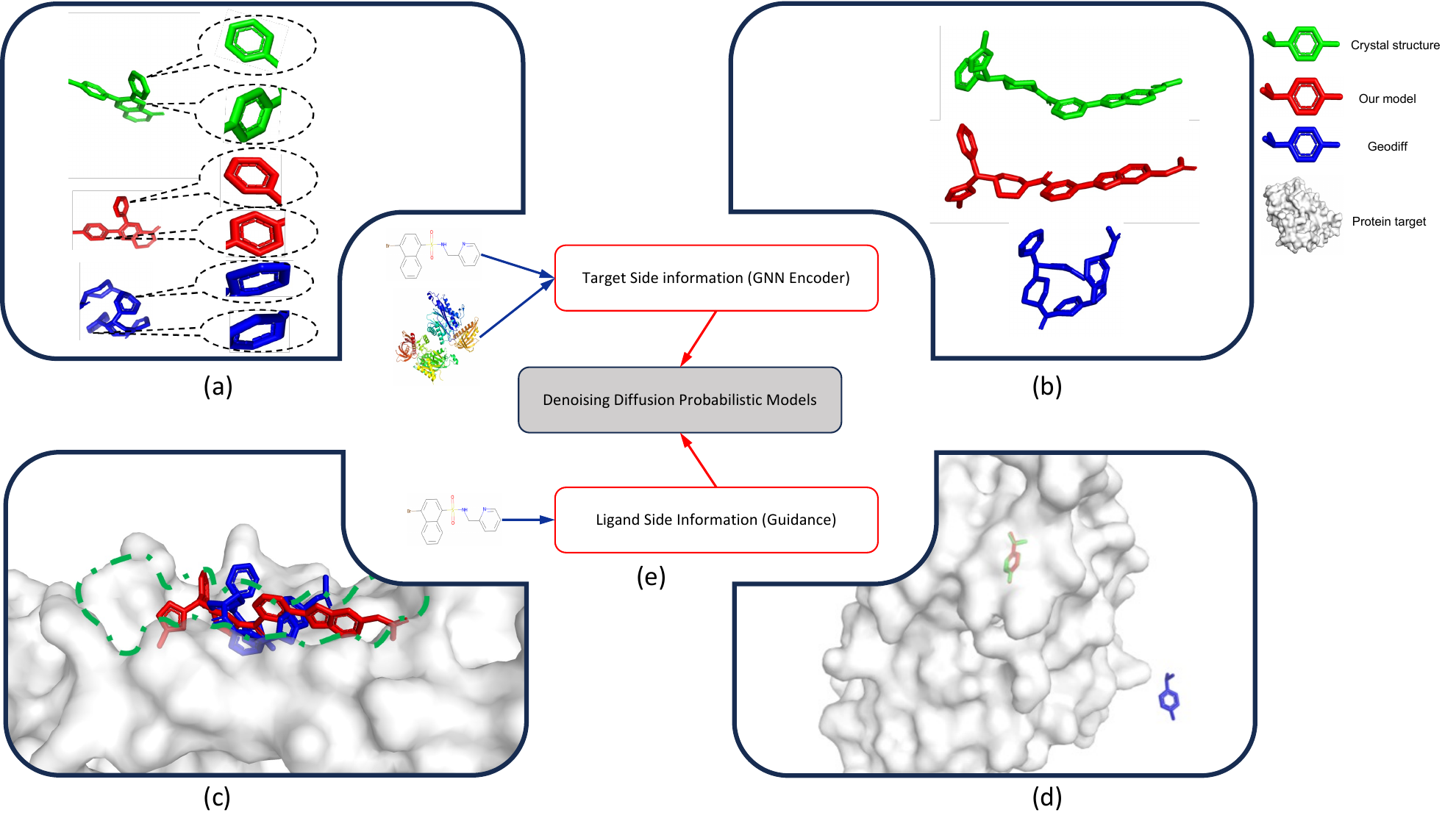}
  \caption{(a)-(d): Comparison between GeoDiff-PDBBind2020 and SIDEGEN. \textcolor{blue}{Blue}: GeoDiff-PDBBind2020, \textcolor{red}{Red}: our, \textcolor{green}{Green}: Crystal structure. (e) is the overview of our model. (a). Biological semantics (e.g. the coplanarity of benzene rings) is omitted in GeoDiff-PDBBind2020; (b). 6cf7. For extra long-range non-covalent interaction, existing work can not catch them. The conformation for GeoDiff-PDBBind2020 is aligned for comparison); (c). 6cf7. The target pocket is circled by the green dot line. The conformation for GeoDiff-PDBBind2020 is aligned for comparison; (d). 6i65. Without alignment, GeoDiff-PDBBind2020 ignores the pocket position and orientation ('pose' of ligands); (e) Overview of our model, To enhance molecular conformation generation using Denoising Diffusion Probabilistic Models, additional target side information is incorporated through a graph neural network-based encoder. Meanwhile, scalar chemical property side information of the ligand is integrated through energy guidance during sampling.}
  \label{fig:intro_compare}
\end{figure}

\section{Introduction}

Drug discovery is a highly time-consuming process, primarily due to the immense search space it entails. It has been estimated to be around $10^{60}$ molecular structures to search \cite{drug_space, MolGenSurvey}. Within the realm of drug design, generating rational ligand conformations from ligand molecular graphs poses a significant challenge. However, deep learning methods have shown promise by enabling the selection and ranking of the most promising candidates. This approach allows experiments to be conducted solely on these candidates, resulting in significant time and cost savings \cite{MLindrug, equibind}. To address this challenge, recent years have witnessed the development of deep learning-based generative models like CVGAE \cite{cvgae}, CONFGF \cite{confgf}, and GeoDiff \cite{GeoDiff}. These models utilize techniques such as variational autoencoders (VAE) and diffusion to generate multiple ligand conformations.\\

However, when deep learning models operate without crucial side information, such as global interactions like target protein chemical and geometric features, ligand-target compound interactions, and local ligand chemical properties, the generated conformations may lack meaningful context for drug design and selection. As depicted in Figure \ref{fig:intro_compare}(c), GeoDiff, for instance, disregards the chemical and geometric information of the target, resulting in drugs that are unsuitable for the intended pocket. While existing methods consider long-range non-covalent interactions within ligands, they still generate chemically invalid conformations illustrated in Figure \ref{fig:intro_compare}(b) due to inadequate constraints. Furthermore, GeoDiff fails to account for the molecular "pose" as it lacks information about the pocket position, as shown in Figure \ref{fig:intro_compare}(d). Moreover, due to the inherent stochasticity in standard Denoising Diffusion Probabilistic Models (DDPM), GeoDiff encounters difficulties in generating conformations with desired semantics, leading to problems like those demonstrated in Figure \ref{fig:intro_compare}(a).
\\

To address these limitations, we propose a side-information-guided diffusion model that constrains the standard sampling progress to align with biological semantics at the local ligand, ligand-target interaction, and compound levels. This model addresses the issues of context preservation and semantic relevance by incorporating crucial side information. It ensures the generation of conformations that respect the chemical and geometric features of the target protein, ligand-target compound interactions, and local ligand chemical properties. By doing so, it enables the generation of ligand conformations that possess meaningful context for drug design and selection. \\

 In SIDEGEN, we tackle the aforementioned challenges through two primary approaches. Firstly, we focus on enhancing the diffusion sampling steps to generate ligand conformations that possess meaningful context and adhere to biological semantics. Ligand molecular graphs not only capture connectivity and node type information but also encapsulate crucial chemical properties such as Self-consistent field (SCF) energy, molecular orbital (HOMO)–lowest unoccupied molecular orbital (LUMO) energy gaps, and Marsili-Gasteiger Partial Charges. To incorporate these properties, we encode them as ligand node features, treating them as significant indicators of chemical and biological relevance. To effectively utilize these properties, with the thought of energy guidance model \cite{guidance, eegsde, EGSDE}, we train additional energy models to predict the scalar chemical properties. By incorporating these energy models, we can guide the diffusion sampling process by leveraging the predicted properties as guidance signals. This enables us to make subtle adjustments to the denoising directions at each step, utilizing invariant energy functions. As a result, the generation process becomes more controlled and directed, leading to the production of ligand conformations that exhibit greater meaningfulness. For instance, as depicted in Figure \ref{fig:intro_compare}(a), this guidance facilitates the preservation of benzene ring coplanarity, which is a crucial structural characteristic in many drugs. \\

 The second key approach in SIDEGEN involves the design of the graph neural network (GNN) to incorporate biological and chemical target information and impose constraints on ligand conformations. To achieve this, we draw inspiration from the concept of message passing in GNNs \cite{gnn} and introduce a feature assembler called ligand-target massage passing block (LTMP). LTMP treats the ligand and compound as two nodes within a graph and performs message passing on a fully connected, directed, and self-cycled graph that involves these two nodes. By extracting the relevant node features, LTMP facilitates the exchange of information between the ligand and compound nodes. This design allows us to incorporate the biological and chemical target information into the ligand conformation generation process, as illustrated in Figure \ref{fig:intro_compare}(c). Consequently, the generated conformations are influenced by and aligned with this crucial side information. To consider long-range interactions both within the ligand and between the ligand and target, we construct ligand-target compound graphs and introduce non-covalent edges based on Euclidean distances. This enables the GNN to capture the long-range interactions within the ligand and between the ligand and target, as depicted in Figure \ref{fig:intro_compare}(b). By incorporating these long-range interactions, the model gains a better understanding of the overall shape and position of ligand conformations, as shown in Figure \ref{fig:intro_compare}(d). This ensures that the generated conformations are not only structurally meaningful but also aligned with the intended position within the target binding site. \\

Overall, SIDEGEN offers a robust solution for ligand conformation generation, leveraging side information and employing a combination of diffusion modeling and graph neural networks. The proposed method contributes to advancing the field of drug discovery by generating ligand conformations that exhibit improved biological relevance and meaningfulness.
In summary, our work makes several contributions to the task of ligand conformation generation: 
\begin{itemize}
    \item We propose a comprehensive diffusion model, which takes into account side information and is guided by biological property energy. This model enables the generation of ligand conformations that possess meaningful biological semantics, enhancing their relevance and usefulness in drug design.
    \item We introduce Side-Information Conditioned Noise Encoder (SICNE): which captures both ligand-target interaction and compound. SICNE incorporates the ligand-target massage passing block (LTMP) feature assembling block, which facilitates the exchange of information between ligand and compound nodes. By constructing ligand-target compounds and considering non-covalent interactions, our model effectively captures the relevant structural and positional information needed for ligand conformation generation.
    \item Experimental results on the PDBBind-2020 dataset demonstrate the effectiveness of SIDEGEN. We observe significant improvements in Aligned RMSD results compared to the baseline, achieving an enhancement of approximately 20\%.

\end{itemize}

\section{Related Work}
\paragraph{Molecular conformation Generation}
Over the past few years, generation models have become increasingly popular in the molecular conformation generation problem. GraphAF \cite{GraphAF} and CVGAE \cite{cvgae}introduced flow-based and VAE-based models, respectively, for molecular coordinates. However, these models failed to address the issue of rot-translation equivariance, which is a crucial consideration in the Euclidean coordinate system. To address this, CGCF \cite{cgcf} and GRAPHDG \cite{graphdg} utilized models on the distance map between atoms rather than 3D coordinates directly. However, these non-end-to-end models require post-processing searching or optimization algorithms to obtain the final 3D coordinates, leading to performance dependency. CONFGF \cite{confgf} tackled this issue by estimating the gradient fields of the log density of atomic coordinates using end-to-end models on denoising score matching methods, but encountered out-of-distribution problems \cite{GeoDiff}. GeoDiff \cite{GeoDiff} and GEOLDM \cite{geoldm} utilized diffusion models on atom space and latent space, respectively. However, all these methods overlooked the importance of target information, which is critical in drug design since different conformations should be designed for different target pockets. Our method addresses this by incorporating both ligand and target information to generate molecular conformations.
% \paragraph{GNN}

\paragraph{Drug-Target Docking Problem}

Drug-target interaction (DTI) problems play a significant role in drug discovery by finding the suitable binding pose of ligand conformations onto some targets \cite{gnina, glide}.
In recent years, graph-based methods have emerged as a promising approach for addressing these problems. EquiBind \cite{equibind} and TANKBind \cite{tankbind} are two such methods that use graph neural networks to predict the coordinates of ligands and identify the binding pocket on the rigid protein. However, these methods are primarily focused on generating a single, optimal binding pose and may not capture the full conformational space of the ligand. Additionally, TANKBind requires further optimization from the ligand-target distance map to the ligand Euclidean coordinates. Furthermore, both DiffDock \cite{DiffDock} and EquiBind require RDKit initialization at the beginning, which involves changing the atom positions by rotating and translating the entire molecule and rotating the torsion angles of the rotatable bonds. This initialization step can be problematic for molecules that cannot be initialized by RDKit \cite{rdkit}, and limits the applicability of these methods to binding-pose conformation generation tasks \cite{MolGenSurvey}. In our method, the initialization is Gaussian noise without any priorities.\\
% \section{Preliminaries}

\section{Methods}

\paragraph{Overview}

In this section, we present the side information conditioned diffusion system in detail, along with its network structures. In Section \ref{sec:problem_definition}, we define the conditioned ligand conformation generation problem. In Section \ref{sec:guided_diffusion}, we provide a high-level formulation for the forward and reverse processes of the diffusion model shown in Figure \ref{fig:diffusion}(b), as well as our proposed ligand chemical property guidance improvement shown in Figure \ref{fig:diffusion}(c) on the existing tasks. We also describe the parameterization of $P_\theta(\mathbf{X}_L|\mathscr{G}_P, \mathscr{G}_L)$, specifically the noise prediction network $\mathbf{s}_{\theta}$ shown in Figure \ref{fig:diffusion}(a) in Section \ref{epsnet}. In Section \ref{sec:equivariance}, we briefly show the normalization and rot-translation invariance of our model. Finally, in Section \ref{training and sampling}, we outline our training and sampling algorithms.

\subsection{Problem Definition} \label{sec:problem_definition}
% \paragraph{Notation}

\begin{table}
    \centering
    \begin{tabular}{c|c}
    \hline
    Notations \\
    \hline
    $\mathscr{G}_L$ & Ligand molecule graph \\
    $\mathscr{G}_P$ & Target graph \\
    $\mathbf{X}_L, \mathbf{X}_P \in \mathbb{R}^3$ & Ligand and target coordinates \\
    $C_L, center_P \in \mathbb{R}^3$ & Ligand and target center \\
    $P_\theta (\mathbf{X}_L|\mathscr{G}_P, \mathscr{G}_L)$ & Parameterized distribution \\
    $j, j'$ & Node index for ligand graphs \\
    $i, i'$ & Node index for target graphs \\
    $m, n$ & Number of nodes in target and ligand \\
    % $d_l, d_p$ & Dimension of ligand and target graph features \\
    $N_L, \mathbf{F}_{L} \in \mathbb{R}^{d_l\times n}$ & Ligand node and node features \\
    $N_P, \mathbf{F}_{P} \in \mathbb{R}^{d_p\times m}$ & Target node and features \\
    $N_C, \mathbf{F}_{C} \in \mathbb{R}^{d_p\times m}$ & Lig-Tar compound node and features \\
    $\mathbf{Z} \in \mathbb{R}^{m\times n \times d}$ & Concat ligand and target feature \\
    $\mathbf{D}_T, \mathbf{D}_L, \mathbf{D}_{inter}$ & Target, ligand, inter pairwise distances \\
    $\mathbf{E}_{ii'}, \mathbf{E}_{jj'}, \mathbf{E}_{ij}$ & Target, ligand, inter edge features \\
    $\mathbf{s}_\theta$ & Parameterized score funtion\\
    $G$ & Energy Guidance \\
    $c$ & Chemical Properties \\
    \hline
    \end{tabular}
    \caption{Notations used in the paper}
    \label{tab:notation}
\end{table}
% \paragraph{Problem definition}
The problem at hand is to generate a conformation that is conditional on the target and ligand molecule graphs. This is defined as a \textit{given target conditional conformation generation} task, where the conditions for the generation task are the target graphs $\mathscr{G}_P$ and the ligand graphs $\mathscr{G}_L$. 

Formally, the objective is to learn a parameterized distribution $P_\theta(\mathbf{X}_L|\mathscr{G}_P, \mathscr{G}_L)$ that approximates the Boltzmann distribution, which represents the probability distribution of conformations for a given ligand molecular graph \cite{Boltzmann}. The learned distribution can then be used to sample i.i.d. conformation coordinates. In other words, given the target and ligand molecule graphs, our goal is to learn a probability distribution that generates conformations consistent with the given conditions. By learning this distribution, we can generate conformations that are more biologically meaningful and relevant for drug design, which can ultimately lead to the discovery of new drugs. The key notations used in this paper are in Table.~\ref{tab:notation}.\\
\subsection{Formulation}
\paragraph{Background of Diffusion Model}
In the forward process, the goal is to get a Markov chain according to a fixed variance schedule $\beta_1,...,\beta_T$ from the actual data distribution $\mathbf{X}_{L_0} \sim q(\mathbf{X}_{L_0})$ to the random Gaussian noise $\mathbf{X}_{L_T} \sim \mathcal{N}(0, \mathbf{I})$ \cite{GeoDiff}:
\begin{equation} \label{eq: forward_kernel}
    q(\mathbf{X}_{L_t}|\mathbf{X}_{L_{t-1}}) = \mathcal{N}(\mathbf{X}_{L_t};\sqrt{1-\beta_t}\mathbf{X}_{L_{t-1}},\beta_t\mathbf{I})
\end{equation}
\begin{equation}\label{eq: forward_from0}
    q(\mathbf{X}_{L_{1:T}}|\mathbf{X}_{L_0}) = \prod_{t=1}^{T}q(\mathbf{X}_{L_t}|\mathbf{X}_{L_{t-1}})
\end{equation}
According to \cite{diffusion}, to simplify the representation of $q(\mathbf{X}_{L_{1:T}}|\mathbf{X}_{L_0})$, let $\alpha_t = 1-\beta_t$ and $\bar{\alpha_t} = \prod_{s=1}^{t}\alpha_s$, then:
\begin{equation} \label{eq: forward}
    q(\mathbf{X}_{L_{1:T}}|\mathbf{X}_{L_0}) = \mathcal{N}(\mathbf{X}_{L_t};\sqrt{\bar{\alpha_t}}\mathbf{X}_{L_{0}},(1-\bar{\alpha_t})\mathbf{I})
\end{equation}
In the reverse process, the goal is to get the conformation at time 0 from an approximate distribution $p_{\theta}$ start from the random Gaussian $\mathbf{X}_{L_T} \sim \mathcal{N}(0, \mathbf{I})$. Formally,
\begin{equation} \label{eq: reverse_kernel}
    p_{\theta}(\mathbf{X}_{L_{t-1}}|\mathbf{X}_{L_{t}}, \mathscr{G}_P, \mathscr{G}_L)
    = \mathcal{N}(\mathbf{X}_{L_{t-1}};\mu_{\theta}(\mathscr{G}_L, \mathscr{G}_P, \mathbf{X}_{L_t}, \tilde{\mathbf{X}}_P, t),\sigma_t\mathbf{I})
\end{equation}
\begin{equation}\label{eq: reverse_T}
   p_{\theta}(\mathbf{X}_{L_{0:T-1}}|\mathbf{X}_{L_T}, \mathscr{G}_P, \mathscr{G}_L)
   = \prod_{t=1}^{T}p_{\theta}(\mathbf{X}_{L_{t-1}}|\mathbf{X}_{L_{t}}, \mathscr{G}_P, \mathscr{G}_L)
\end{equation}
where $\tilde{\mathbf{X}}_P$ is the normalized target position calculated in Eq.~\ref{norm}. Here $\mu_{\theta}$ and $\sigma_t$ are the mean and standard deviation of the approximate distribution as follows:
\begin{equation} \label{eq: mu}
   \mu_{\theta} = \frac{1}{\sqrt{\alpha_t}}(\mathbf{X}_{L_{t}} - \beta_t\frac{\mathbf{s}_{\theta}}{\sqrt{1-\bar{\alpha_t}}})
\end{equation}
\begin{equation} \label{eq:sigma}
   \sigma_t = \beta_t\frac{1-\bar{\alpha}_{t-1}}{1-\alpha_t}
\end{equation}
where $\mathbf{s}_{\theta}$ is the parameterized noise calculated by the neural network. \\

\paragraph{Chemical Property-Energy Guided Diffusion} \label{sec:guided_diffusion}
\begin{figure}[!hb]
  \centering
  \includegraphics[width=0.98\columnwidth]{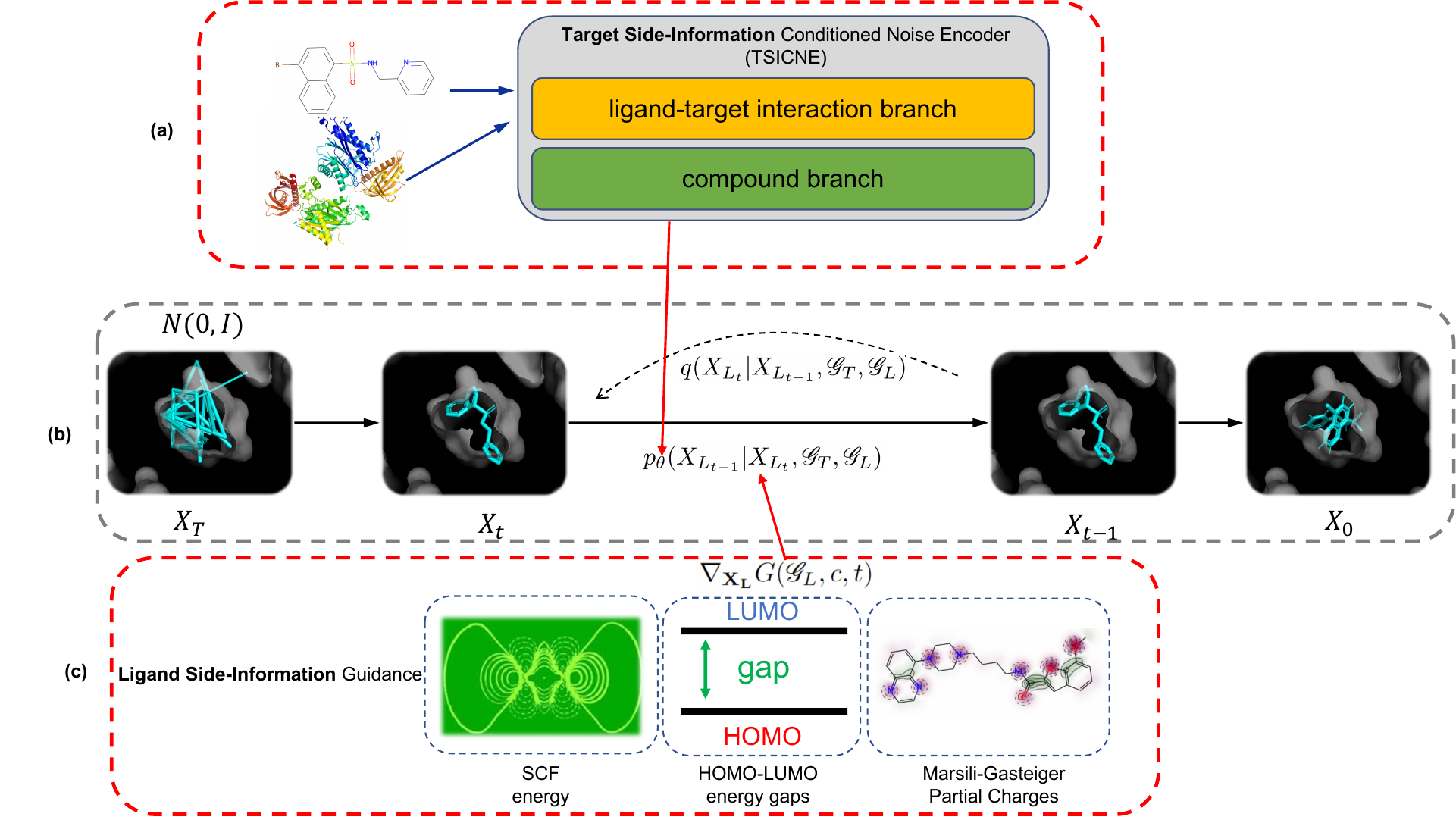}
  \caption{Overview of SIDEGEN Ligand molecular graph and target point cloud are regarded as side information to catch the ligand shape and 'pose' in TSICNE in both ligand-target interaction and compound manners as shown in (a). Chemical properties (SCF energy, HOMO-LUMO energy gaps, and Marsili-Gasteiger Partial Charges) are utilized to make subtle adjustments for ligand conformation sampling as shown in (c). (b) is the standard DDPM progress used in GeoDiff.}
  \label{fig:diffusion}
\end{figure}

% \paragraph{Diffusion overview}
As illustrated in Figure \ref{fig:diffusion}, the diffusion generation process can be viewed as a two-step process: a forward process and a reverse process. In the forward process, noise is added to the samples that are drawn from the ligand Boltzmann distribution, $\mathbf{X}_{L_0} \sim q(\mathbf{X}_{L_0})$. This process generates a sequence of samples that approximate the target distribution. In the reverse process, a denoising process is used to obtain the approximate distribution, $p_{\theta}$, from standard Gaussian distributions. \\

To generate molecular conformations from Gaussian noise, we need to reverse the diffusion process and the process can be represented by the stochastic differential equation (SDE) shown in Eq.~\ref{eq:sde} \cite{ddpm, GeoDiff, EGSDE}, 
\begin{equation}\label{eq:sde}
   d\mathbf{X}_L = [f(N_L, c, t)\mathbf{X}_Ldt + g(t)^2\mathbf{s}_t(\mathbf{X}_L, N_L, c, t)dt] 
   + g(t)\overline{\mathbf{\omega_{\mathbf{X}_L}}}
\end{equation}
where $f(t)$ and $s(t)$ are two scalar functions while $\mathbf{s}_t(\mathbf{X}_L, N_L, c, t)$ is the score function, which can be parameterized by the noise prediction network $\mathbf{s}_{\theta}$ in Section \ref{epsnet}. To train the noise prediction network, we minimize the MSE loss in Eq.~\ref{eq:mse_loss}.\\

\begin{equation} \label{eq:mse_loss}
    \mathcal{L} = \mathbb{E}[\| \mathbf{s}_{\theta} - \mathbf{s} \|^2]
\end{equation}

% \paragraph{Guidance}
Different from GeoDiff \cite{GeoDiff} shown in Figure \ref{fig:diffusion}(b), we incorporate chemical properties to guide the sampling process, thereby preserving the chemical semantics of the ligands as shown in Figure \ref{fig:diffusion}(c).
Given a molecular Simplified Molecular Input Line Entry System (SMILES) \cite{SMILES}, we can easily calculate various chemical properties, such as Self-consistent field (SCF) energy, molecular orbital (HOMO)–lowest unoccupied molecular orbital (LUMO) energy gaps, and Marsili-Gasteiger Partial Charges, using chemical tools such as Psi4 \cite{Psi411}. Following the approach outlined in \cite{EGSDE}, we can guide the SDE used by GeoDiff \cite{GeoDiff} shown in Eq.~\ref{eq:sde}  with the gradient of an energy function $\nabla_{\mathbf{X}_L}G(\mathscr{G}_L, c, t)$, where $c$ and $t$ denote the chemical properties and the time step, respectively.

Formally, the reverse SDE can be described in Eq.~\ref{eq:guide_sde}. Here, $f(N_L, c, t)$ and $g(t)$ are scalar functions, $\overline{\mathbf{\omega_{\mathbf{X}_L}}}$ is the reverse standard Wiener process, and $\mathbf{s}_t(\mathbf{X}_L, N_L, c, t)$ is the score estimated by the network in Section \ref{epsnet}. We also introduce energy-guidance models, $G_{energy}$, $G_{gap}$, and $G_{charge}$ shown in Eq.~\ref{eq:guidance_models}, which are trained to predict the chemical properties mentioned above. Additionally, $\lambda_{energy}$, $\lambda_{gap}$, and $\lambda_{charge}$ are scalar weights on the guidance.\\

\begin{equation}\label{eq:guide_sde}
   d\mathbf{X}_L = [f(N_L, c, t)\mathbf{X}_Ldt + g(t)^2(\mathbf{s}_t(\mathbf{X}_L, N_L, c, t) 
   + \textcolor{red}{\omega(\mathscr{G}_L, c, t)})dt]
   + g(t)\overline{\mathbf{\omega_{\mathbf{X}_L}}} 
\end{equation}
where 
\begin{equation}\label{eq:guidance_models}
    \omega = \lambda_{energy}\nabla_{\mathbf{X}_L}G_{energy}(\mathscr{G}_L, c, t) 
   + \lambda_{gap}\nabla_{\mathbf{X}_L}G_{gap}(\mathscr{G}_L, c, t) 
   + \lambda_{charge}\nabla_{\mathbf{X}_L}G_{charge}(\mathscr{G}_L, c, t))dt]
\end{equation}
   
Following \cite{GeoDiff, EGSDE}, samples can be sampled from the approximate Gaussian distribution from time step T to 1 with $\mu_t$ and $\sigma_t$ defined in Eq.~\ref{eq:guide_mu} and Eq.~\ref{eq:sigma}. The final coordinates can be sampled from $p(\mathbf{X}_L|\mathbf{X}_{L_0})$ with $\mu_0$ and $\sigma_0$ defined in Eq.~\ref{eq:guide_init}. \\

\begin{equation} \label{eq:guide_mu}
   \mu_t = \frac{1}{\sqrt{\bar{\alpha}_t}}(\mathbf{X}_{L_t} - \frac{1-\bar{\alpha}_t}{\sqrt{1-\bar{\alpha}}_t}\mathbf{s}_{\theta}) + \lambda_{prop}\psi_{prop}
\end{equation}
where $\lambda_{prop}$ is the weight of chemical properties including Self-consistent field (SCF) energy, molecular orbital (HOMO)–lowest unoccupied molecular orbital (LUMO) energy gaps, and Marsili-Gasteiger Partial Charges, $\psi_{prop}$ in Eq.~\ref{eq:prop} denotes the guidance model predicting the properties above.
\begin{equation} \label{eq:prop}
   \psi_{prop} = \sqrt{1 - (\frac{\bar{\alpha}_{t-1}}{\bar{\alpha}_t})^2}\nabla_{\theta'}\|G_{prop_{\theta'}} - c_{prop}\|^2
\end{equation}
\begin{equation} \label{eq:guide_init}
   \mu_0 = \sqrt{\frac{1}{\bar{\alpha}}}(\mathbf{X}_{L_1} - \sqrt{\frac{1}{1+\frac{\bar{\alpha}}{1 - \bar{\alpha}}}}\mathbf{s}_{\theta}), 
   \sigma_0 = \sqrt{\frac{1 - \bar{\alpha}_0}{\bar{\alpha}_0}}
\end{equation}

To train the guidance model $\psi_{prop}$, we use an Equivariant Graph Convolution Layer (EGCL) \cite{EGNN, EGNN_gene}-based model, which is rotationally invariant. This is because the operations on the coordinate space are linear, and the features are scalars, which are always invariant. The details for the guidance network are provided in Appendix~\ref{appendix:guidance}, while the rotational invariant proof is given in Appendix~\ref{appen:se3 proof}. By incorporating these predicted properties into the reverse SDE, we can generate conformations that are more biologically meaningful and relevant for drug discovery. \\

\subsection{Target Side-Information Conditioned Noise Encoder (TSICNE)} \label{epsnet}

In this section, we provide a detailed parameterized encoder shown in Figure \ref{fig:diffusion}(c) for $P_\theta(\mathbf{X}_L|\mathscr{G}_P, \mathscr{G}_L)$. The encoder $\mathbf{s}_{\theta}$ is designed to approximate estimate the score function $\mathbf{s}_t(\mathbf{X}_L, N_L, c, t)$ shown in Eq.~\ref{eq:sde}. The input to the network consists of the ligand graphs $\mathscr{G}_L$ and target graphs $\mathscr{G}_P$, as described in Appendix~\ref{appen:graph repr}.

\begin{figure}[!hb]
  \centering
  \includegraphics[width=0.95\columnwidth]{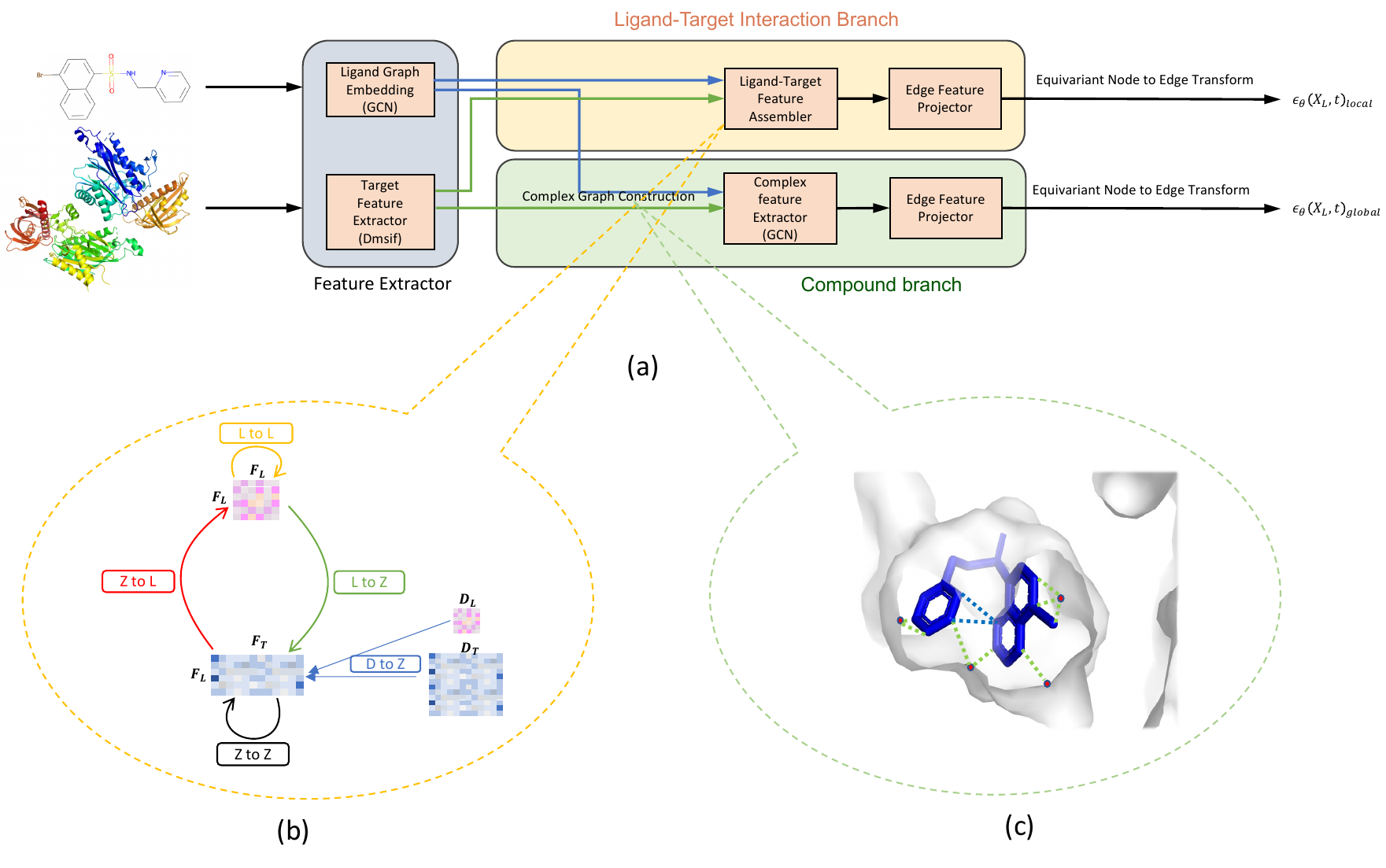}
  \caption{(a). Side-Information Conditioned Noise Encoder (SICNE) overview, consists of the ligand-target interaction branch and compound branch; (b). Overview of ligand-target massage passing (LTMP) block: $\mathbf{F}_L$ and $\mathbf{F}_P$ denote the ligand and target node feature, respectively. $\mathbf{D}_L$ and $\mathbf{D}_T$ denote the ligand and target pairwise distances, respectively. The message passing among ligand node feature, ligand-target assembled node feature, and pairwise distances; (c). Ligand-target compound edge construction: the {\color{red} red} dots represent the protein surface nodes. The edges between ligand and target graphs encode the \textit{inter-interaction} expressing as {\color{green} green} dashed lines and the new edges inside ligand graphs bring the long-range effects for the non-covalent nodes into consideration expressing as {\color{blue} blue} dashed lines.}
  \label{fig:encoder}
\end{figure}

The encoder network SICNE is a crucial element of the system described and is composed of two distinct branches: the ligand-target interaction and compound encoders, which are illustrated in Figure \ref{fig:encoder}(a).

The ligand-target interaction encoder extracts node features from both the ligand and target inputs, which are subsequently merged using LTMP, a feature assembling block. Conversely, the compound branch encoder constructs a compound graph by merging the ligand and target graphs, and applies a graph neural network to extract compound features.

Both the ligand-target interaction and compound branches produce edge and node features that are combined. To ensure roto-translate invariance across both branches, an equivariant transformation is performed on the projected edge features. This procedure enables the model to effectively learn from both ligand-target interaction and compound features, which in turn enables it to identify meaningful interactions between the ligand and target molecules. \\

\paragraph{ligand-target interaction branch}
In the ligand-target interaction branch, the first step is to extract features for both ligand and target.
For the target graph, we follow the Dmasif \cite{dmasif} feature extractor and embed the target as a point cloud graph with node features consisting of two components: chemical features and geometric features. To capture the shape of the pocket surface, we select point clouds close to the surface as the nodes of the target graphs using the signed distance function (SDF) in Eq.\ref{SDF}. This is because the surface of the target determines most of the properties for the generated ligand conformations \cite{sdf1,sdf2,sdf3,sdf4}. Here, $\textbf{a}_j$ denotes the protein atoms, $\mathbf{N_P}$ denotes the selected point clouds (nodes of the target graph), $\sigma$ is the experimental atom radius for $\textbf{a}_j$, and w is the averaged atom radius. Additional details are provided in Appendix\ref{appen:graph repr}.\\
\begin{equation}\label{SDF}
    \textrm{SDF}(\mathbf{N_P}) = - \textit{w} \cdot \log \sum_{j=1}^m \exp ( - \| \mathbf{N_P} - \textbf{a}_j\| / \sigma)
\end{equation}
The ligand graphs are represented by molecular graphs with edges in h-tops (h=3), as described in Appendix~\ref{appen:graph repr}. The ligand-target interaction graph encodes the interaction of nodes with chemical bonds, constructing the local structure, such as the ionic, polar covalent, and electric interactions.

While the molecular graph provides the above through strong chemical interactions, using only chemical bonds as edges ignore the long-range connections for nodes without chemical bonds but located near each other in Euclidean space. Additionally, without absolute coordinate information for the target graphs, the binding pose corresponding to the target is neglected.

To overcome the limitations of previous approaches, we have integrated non-covalent interactions into our methodology. Specifically, when the Euclidean distance between two ligand nodes is less than a designated threshold, we create pseudo edges between them. Additionally, the distance between these nodes is encoded as part of the edge features, allowing our approach to incorporate additional information about the spatial relationships between ligand nodes. \\

In our approach, we use a Graph Isomorphism Network (GIN) for the ligand-target interaction branch as the ligand feature extractor in equations \ref{eq: feat_extractor1} and \ref{eq: feat_extractor2}. For the target point cloud graph, we follow the approach of \cite{dmasif} and extract the geometric and chemical features in equation \ref{eq: feat_extractor3}. Here, $f_{chem_i}^{l}$ and $f_{geom_{ii'}^{l}}$ denote the chemical and geometric features for the target nodes, respectively. $\Phi_{m_{local}}$ and $\Phi_{h_{local}}$ denotes the parameterized ligand-target interaction networks. $\theta_{m_{local}}$ and $\theta_{h_{local}}$ denotes the parameters in the ligand-target interaction branch.
\begin{equation} \label{eq: feat_extractor1}
    \mathbf{m_{jj'}} = \Phi_{m_{local}}(\mathbf{F}_{L_j}^l, \mathbf{F}_{L_j'}^l, D_{jj'}, \mathbf{E}_{jj'};\theta_{m_{local}})
\end{equation}
\begin{equation} \label{eq: feat_extractor2}
    \mathbf{F}_{L_j}^{l+1} = \Phi_{h_{local}}(\mathbf{F}_{L_j}^l, \sum_{j' \in N(j)}\mathbf{m_{jj'}};\theta_{h_{local}})
\end{equation}
\begin{equation} \label{eq: feat_extractor3}
    \mathbf{F}_{P_{i}}^{l+1} = \Phi_p(f_{chem_i}^{l}, f_{geom_ii'}^{l})
\end{equation}
    
While our approach uses a combination of GIN and geometric and chemical feature extraction to capture both ligand-target interaction and compound features of the input graphs, using all the sampled point clouds can result in a feature assembler that is computationally expensive. Additionally, dense target features may be redundant when the features are already extracted.

To address these issues, we use Fastest Point Sampling (FPS) \cite{fps1,fps2} to downsample the target point clouds after features are extracted. This enables us to reduce the computational cost of the feature assembler while still preserving the relevant information needed for generating biologically meaningful conformations. \\

Once the features have been extracted from the ligand and target graphs, the next step is to facilitate communication between the two sets of features. To accomplish this, we have devised a feature assembling block called ligand-target massage passing block (LTMP). This block is specifically designed to transfer the information from the ligand and target graphs, enabling them to interact and exchange information. Inspired by the message passing thought, we regard the ligand node features $\mathbf{F}_{L}$ and the concatenated ligand-target node features $\mathbf{Z} \in \mathbb{R}^{m\times n \times d}$ as two nodes of a directed, self-looped fully connected graph. However, using only the node feature assembler may miss some internal information for the ligand and target. To address this limitation, we also add the 'D to Z' block, inspired by \cite{tankbind}, to update the concatenated feature by the ligand-ligand distance $\mathbf{D}_L$ and the target-target distance $\mathbf{D}_T$.

To pass messages between the two nodes, we design five sub-blocks to update the graph in each layer, as shown in Figure \ref{fig:encoder}(b), to cover each of the edges, and finally output $\mathbf{Z}$ after several layers. The detailed sub-block design is provided in Appendix\ref{appendix:LTMP}.
\begin{equation}\label{eq: LTMP}
   \mathbf{Z} = LTMP(\mathbf{F}_L, \mathbf{Z})
\end{equation}
where $\mathbf{Z} = Concat(\mathbf{F}_L, \mathbf{F}_P) \in \mathbb{R}^{m\times n \times d}$. \\
In our approach, the targets are regarded as fixed and rigid, and the partitions to update belong to the ligand graphs only. Therefore, we transfer the concatenated node feature to the ligand nodes by using average pooling. After that, we obtain the output feature $\mathbf{F}_{L_{out_{local}}}$ by using an MLP on the concatenation of the ligand node and edge features $\mathbf{E}_{local}$, as shown in Eq.~\ref{eq: eq-out}.
\begin{equation}\label{eq: eq-out}
    \mathbf{F}_{L_{out_{local}}}
    = MLP(Concat(AdaptiveAveragePool(\mathbf{Z}), \mathbf{E}_{local}))
\end{equation}

\paragraph{Compound branch}
To better interpret the intra-ligand long-range interaction and the ligand-target 'inter-graph interaction, which determines the binding pose, we construct the ligand-target compound graph with node features the same as the ligand-target interaction ligand and target graphs. We add edges between nodes of both the ligand and target within some distance cutoffs, as shown in Figure \ref{fig:encoder}(c). The edges between the ligand and target graphs encode the 'inter-interaction' expressed as green dashed lines, while the new edges inside the ligand graphs bring the long-range effects for the non-covalent nodes into consideration expressed as blue dashed lines. The target graph is considered a condition and remains fixed during the diffusion process; therefore, the edges inside the target are ignored. Ablation studies show the effectiveness of the compound edges in Section\ref{sec:exp}. \\

After using the target feature extractor in Eq.\ref{eq: feat_extractor3}, we construct the ligand-target compound graph by adding edges between nodes of both the ligand and target within some distance cutoffs. We use SchNet \cite{schnet} as the compound graph feature extractor for message passing in Eq.\ref{eq: complex_gnnm} and \ref{eq: complex_gnnh}. The output node and edge features are concatenated and projected to obtain the edge noise score in Eq.~\ref{eq: complex_out}. Here, $\Phi_{m_{global}}$ and $\Phi_{h_{global}}$ denotes the parameterized compound branch network. $\theta_{m_{global}}$ and $\theta_{h_{global}}$ denotes the parameters in the compound branch.
\begin{equation} \label{eq: complex_gnnm}
    \mathbf{m}_{C_{jy}} = \Phi_{m_{global}}(\mathbf{F}_{C_{j}^l}, \mathbf{F}_{C_{y}}^l, \mathbf{D}_{jy}, \mathbf{E}_{jy};\theta_{m_{global}})
\end{equation}
\begin{equation} \label{eq: complex_gnnh}
    \mathbf{F}_{C_{j}}^{l+1} = \Phi_{h_{global}}(\mathbf{F}_{C_{j}}^l, \sum_{y \in N(j)}\mathbf{m}_{C_{jy}};\theta_{h_{global}})
\end{equation}
\begin{equation} \label{eq: complex_out}
     \mathbf{F}_{L_{out_{global}}} = MLP(\mathbf{F}_{C}^L, \mathbf{E}_{global}))
\end{equation}
Where $y$ denotes the nodes in the ligand-target compound graph, $\mathbf{F}_{L_{out_{global}}}$ is the output feature for the compound branch with the compound edges $\mathbf{E}_{global}$.

% \subsubsection{Feature Assembler} \label{feature assembler}
% In this section, we introduce LTMP as the node feature assembler block for the local branch to transfer the information from the targets into the ligands. 
\subsubsection{Edge-to-node Equivarance Transform} \label{eq-transform}
Once the features have been extracted, we project the node features onto the edges and concatenate the resulting features to obtain the projected edge features. This step is crucial for enabling the model to capture the relationships between nodes and edges, as it allows the edge features to incorporate information about the nodes that they connect.
To ensure rotational invariance, we represent the node position noises using the weighted sum of edge features connecting the node, similar to \cite{GeoDiff}. We use a parameterized score $\mathbf{s}_{\theta}$, as expressed in Eq.~\ref{eq: eq-trans}.
\begin{equation}\label{eq: eq-trans}
    \mathbf{s}_{\theta} = \sum_{j' \in N(j)}dir_{jj'}\mathbf{F}_{L_{out_{jj'}}}
\end{equation}
where $dir_{jj'})$ denotes the unit director of the vector between the coordinates of two nodes, calculated as $dir_{jj'} = \frac{1}{D_{jj'}}(\mathbf{X}_{input_j} - \mathbf{X}_{input_j'})$.

\subsection{Rot-translation invariant} \label{sec:equivariance}
\paragraph{Normalization}
Different from GeoDiff \cite{GeoDiff}, we first normalize all the coordinates to make the scalar of small and large compounds consistent. After the normalization, our model is still rot-translation invariant as the transformation is linear.
To use the standard DDPM sampling process, we first normalize the ligand and target so that their coordinates have the same value range as the standard Gaussian noise in Eq.~\ref{norm}. Here, $var_P$ is the maximum of the variance of the XYZ coordinates for the target, calculated as $var_P = max(var_{P_X}, var_{P_Y}, var_{P_Z})$. This normalization ensures that the value range of the ligand and target coordinates are the same, which is necessary for the diffusion process.

After the sampling process, we transfer the generated conformations back to the original coordinates using the recorded mean and variance, as shown in Eq.~\ref{norm_back}. The targets are considered fixed and rigid, with their centers and variances treated as scalars. Therefore, the normalization transforms for the ligands are rot-translate invariant. We provide detailed proofs of the rot-translate invariance with normalization in Appendix \ref{appen:se3 proof}.\\

\subsection{Training and Sampling} \label{training and sampling}
To ensure that the value ranges of the target and ligand node coordinates remain the same as the noises, which are sampled from standard normal distributions, we normalize the coordinates before taking gradient descent steps on the Epsilon network to train the noise score $\mathbf{s}_{\theta}$ using the loss in Eq.\ref{eq:mse_loss}. The Pseudo code for training is shown in Algorithm.\ref{alg:training}. \\

For the reverse process for sampling, we follow the standard DDPM algorithm with energy guidance on the chemical properties, as shown in Eq.\ref{eq:guide_sde}. After finishing all sampling steps, we transfer the coordinates value range back to the initial coordinates, as shown in the last line of Algorithm.\ref{alg:sampling}. \\

As described in Section~\ref{sec:guided_diffusion}, the energy guidance is defined as the gradient of the L2 norm of the difference between predicted and ground truth chemical features. The training process for the energy guidance is shown in Algorithm.~\ref{alg:train_guidance}.

\begin{algorithm}
\caption{Generation Model Training}
\begin{algorithmic}[1]
\Require $\mathscr{G}_L, \mathscr{G}_P, \mathbf{X}_{L_t}, c, \mathbf{X}_P$
\Repeat
\State $\mathbf{X}_{L_0} \sim q(\mathbf{X}_{L_0})$
\State $\tilde{\mathbf{X}}_{L_0} = \frac{\mathbf{X}_{L_0} - center_P}{\sqrt{var_P}}$ \Comment{Normalize ligand coordinates}
\State $\tilde{\mathbf{X}}_P = \frac{\mathbf{X}_P - center_P}{\sqrt{var_P}}$ \Comment{Normalize target coordinates}
\State $\mathbf{s} \sim \mathcal{N}(0, \mathbf{I})$
\State $\tilde{\mathbf{X}}_{L_t} = \sqrt{\bar{\alpha}_t}\tilde{\mathbf{X}}_{L_0} + \sqrt{1 - \bar{\alpha}_t}\mathbf{s}$ \Comment{Perturb ligand coordinates}
\State $\mathbf{s}_{\theta} = \Phi_{\theta}(\mathscr{G}_L, \mathscr{G}_P, \tilde{\mathbf{X}}_{L_t}, \tilde{\mathbf{X}}_P, c, t)$
\State Take gradient descent step on $$\nabla_{\theta}\| \mathbf{s}_{\theta} - \mathbf{s} \|^2$$ \Comment{Loss function defined in Eq.~\ref{eq:mse_loss}}
\Until{converged} 
\end{algorithmic}
\label{alg:training}
\end{algorithm}

\begin{algorithm}
\caption{Sampling}
\begin{algorithmic}[1]
\Require $\mathscr{G}_L, \mathscr{G}_P, \mathbf{X}_P, c$
\Ensure $\mathbf{X}_{L_0}$
\State $\tilde{\mathbf{X}}_P = \frac{\mathbf{X}_P - center_P}{\sqrt{var_P}}$ \Comment{Normalize target coordinates}
\State $\tilde{\mathbf{X}}_{L_T} \sim \mathcal{N}(0, \mathbf{I})$ \Comment{Random initial ligand coordinates}
\For{$t = T,...,1$}
    \State $\mathbf{z} \sim \mathcal{N}(0, \mathbf{I})$ if $t > 1$, else $\mathbf{z} = 0$
    \State $\mathbf{s}_{\theta} = \Phi_{\theta}(\mathscr{G}_L, \mathscr{G}_P, \tilde{\mathbf{X}}_{L_t}, \tilde{\mathbf{X}}_P, c, t)$
    \State Calculate $\mu_t$ and $\sigma_t$ from Eq.~\ref{eq:guide_mu} and Eq.~\ref{eq:sigma}
    \State $\tilde{\mathbf{X}}_{L_T} = \mu_t + \sigma_t\mathbf{z}$ \Comment{Update ligand coordinates by DDPM with guidance}
    \State $\tilde{\mathbf{X}}_{L_T} = \tilde{\mathbf{X}}_{L_T} - Center(\tilde{\mathbf{X}}_{L_T})$ \Comment{Take CoM}
\EndFor
\State Calculate $\mu_0$ and $\sigma_0$ from Eq.~\ref{eq:guide_init}
\State Sample $\tilde{\mathbf{X}}_{L_0}$ from $\mathcal{N}(\mu_0, \sigma_0)$ \Comment{Sample final coordinates}

\State $\mathbf{X}_{L_0} = \tilde{\mathbf{X}}_{L_0} * \sqrt{var_P} + center_P$ \\
\Comment{Transfer the coordinates back to the initial value range}
% \State \Return $\mathbf{X}_{L_0}$
\end{algorithmic}
\label{alg:sampling}
\end{algorithm}

\begin{algorithm}
\caption{Energy Guidance Model Training}
\begin{algorithmic}[1]
\Require $\mathscr{G}_L, \mathbf{X}_{L_t}, c$
\Repeat
\State $\mathbf{X}_{L_0} \sim q(\mathbf{X}_{L_0})$
\State $\tilde{\mathbf{X}}_{L_0} = \frac{\mathbf{X}_{L_0} - center_P}{\sqrt{var_P}}$ \Comment{Normalize ligand coordinates}

\State $\mathbf{s} \sim \mathcal{N}(0, \mathbf{I})$
\State $\tilde{\mathbf{X}}_{L_t} = \sqrt{\bar{\alpha}_t}\tilde{\mathbf{X}}_{L_0} + \sqrt{1 - \bar{\alpha}_t}\mathbf{s}$ \Comment{Perturb ligand coordinates}
\State $c_{pred} = G_{\theta'}(\mathscr{G}_L, c, t)$ \Comment{Predict chemical features}
\State Take gradient descent step on $$\nabla_{\theta'}| c_{pred} - c_{prop}|$$ \Comment{Loss function defined in Eq.~\ref{eq:guidance_loss}}
\Until{converged}
\end{algorithmic}
\label{alg:train_guidance}
\end{algorithm}

\section{Experiments} \label{sec:exp}
\subsection{Dataset}

We used PDBBind-2020 for both training and sampling in this work. Following the same data splitting strategy as \cite{tankbind} and removing data with atoms outside the 32 atom types or data that cannot be processed by Psi4 or RDKit for property calculation, we obtained 13,412, 1,172, and 337 pairs of compounds in the training, validation, and test sets, respectively. The test set does not contain any data that appear in or are similar to the training or validation sets.

Unlike traditional ligand conformation generation datasets such as GEOM \cite{geom}, which contain no target data, PDBBind contains both ligand and target data, but they have a one-to-one correspondence. This enables us to effectively capture both intra-ligand long-range interactions and ligand-target 'inter-graph' interactions, as described in Section.\ref{epsnet}. \\

\subsection{Experiment Setting}

We used Adam \cite{adam} as the optimizer for both the diffusion and energy guidance models. The diffusion model was trained with 5000 steps for inference in the aligned RMSD experiment and 1000 steps for the RMSD experiments. It took around two days on eight Tesla A100 GPUs to train for 80 epochs.

During sampling, we added compound information only when $\sigma < 0.5$ for ligands with more than 50 atoms (i.e., large ligands) and when $\sigma < 3.4192$ for those with fewer than 50 atoms (i.e., small ligands). For the pseudo-edge threshold, we used $8 \mathring{\textrm{A}}$ as the intra-edge threshold and $2.8 \mathring{\textrm{A}}$ as the inter-edge threshold. Experimentally, atoms within $8 \mathring{\textrm{A}}$ have non-covalent interactions inside a molecule. We chose the inter-edge threshold by first calculating the fraction of the number of atoms in the ligand and pocket, which was $7.08\%$. Then, we chose the $7.08\%$ quantile of the pairwise distances, which was $2.8 \mathring{\textrm{A}}$.
The experiments settings for the chemical property energy model are in Appendix~\ref{appendix:guidance}.
\paragraph{Evaluation Metric}
We evaluate the generation quality in two aspects: similarity to the crystal conformations, which is evaluated by the aligned RMSD in Eq.\ref{eq: aligned_RMSD}, and binding poses, which are evaluated by the Ligand RMSD in Eq.\ref{eq: RMSD}.
For two conformations $\mathbf{X} \in \mathbb{R}^{n\times 3}$ and $\hat{\mathbf{X}} \in \mathbb{R}^{n\times 3}$, the Room-Mean-Square Deviation (RMSD) between them can be written as:
\begin{equation} \label{eq: RMSD}
    RMSD(\mathbf{X}, \hat{\mathbf{X}}) = (\frac{1}{n}\sum_{j=1}^{n}\| \mathbf{X}_j, \hat{\mathbf{X}_j} \|^2)^{\frac{1}{2}}
\end{equation}
If the pose is ignored, with $R_g$ denoting the rotation in SE(3) group, the alignment of two conformations can be evaluated by the Kabsch-aligned RMSD:
\begin{equation} \label{eq: aligned_RMSD}
    RMSDAlign(\mathbf{X}, \hat{\mathbf{X}}) = \mathop{\arg\min}_{\mathbf{X'} \in R_g \hat{\mathbf{X}}}RMSD(\mathbf{X}, \mathbf{X'})
\end{equation}

\subsection{Results on Aligned RMSD}
In this section, we compare the average of five generated conformations and compare them with baseline models.  \\

ligand conformation generation method (GeoDiff \cite{GeoDiff}) and the docking method (TANKBind \cite{tankbind}). To compare the structures fairly, we used the same training set as TANKBind (PDBBind-2020) and retrained the GeoDiff model on the same dataset. The performance of the original weights given by GeoDiff (trained on GEOM-QM9 \cite{qm9} and GEOM-Drugs \cite{drugs} datasets) was worse, and the results are in Appendix~\ref{app: more-results}. Here, we use GeoDiff-PDBBind to denote the GeoDiff model retrained on the PDBBind dataset.\\

The quality of the generated conformations can be evaluated by the aligned RMSD defined in Eq.\ref{eq: aligned_RMSD}. As shown in Table\ref{tab:aligned_rmsd}, without any other optimization, our method reduced the median of aligned RMSD by $20\%$ compared to GeoDiff-PDBBind and $7.6\%$ compared to TANKBind. With a simple force field optimization \cite{ff}, our method can reduce the value by $20\%$ compared to GeoDiff-PDBBind and $10\%$ compared to TANKBind.
%\jm{ranking and selection to be added after experiments}
\begin{table}
    \centering
    \begin{tabular}{c|cccc}
    \hline
    \multirow{2}*{Models}  & \multicolumn{4}{c}{Aligned RMSD($\mathring{\textrm{A}}$)$\downarrow$} \\
     \cline{2-5}
    ~ & mean & 25th & 50th & 75th \\
    \hline
    GeoDiff-PDBBind & 2.79    & 1.61      & 2.47     & 3.58    \\
    TANKBind  & 2.61     & 1.43  & 2.20    & 3.15 \\
    \hline
    SIDEGEN & 2.609  & 1.417  & 2.033  & 3.09   \\
    SIDEGEN + FF & \textbf{2.36} & \textbf{1.335}  & \textbf{1.98}  & \textbf{2.85} \\
    \hline
    \end{tabular}
    \caption{RMSD after alignment by Kabsch algorithm on PDBBind-2020(filtered)}
    \label{tab:aligned_rmsd}
\end{table}

% \paragraph{Visualization with pocket}
\subsection{Ablation study for different structures}
In this section, we assessed the effects of the intra-ligand long-range connection, inter-edges connection between the ligand and target, the LTMP node feature assembler, and guidance through ablation studies. \\

\begin{figure}[ht]
  \centering
  \includegraphics[width=0.95\columnwidth]{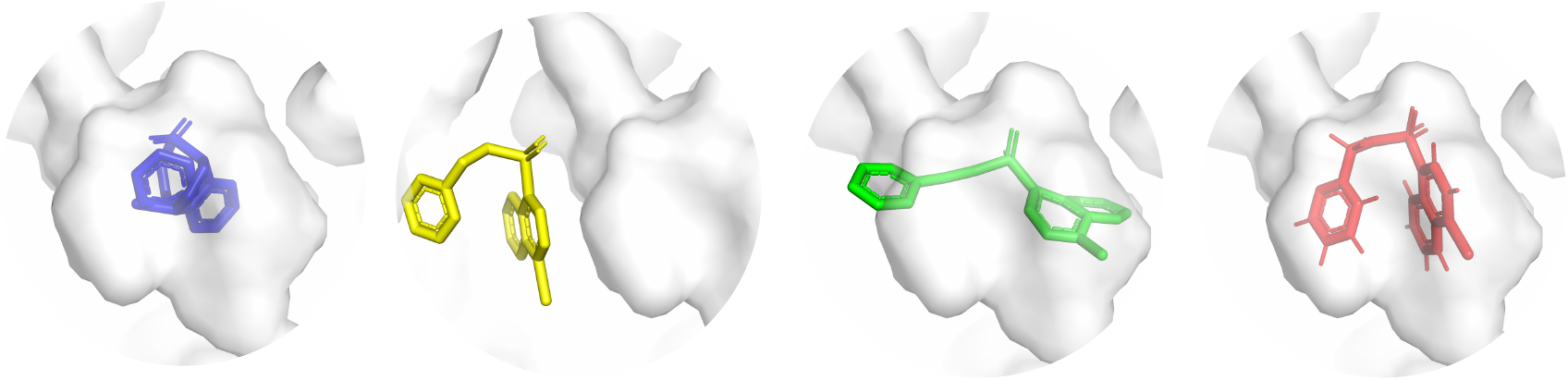}
  \caption{Ablation study for the effect of the intra-ligand long-range connection, the inter-edges connection between ligand and LTMP. The blue ligand is generated without intra-ligand long-range edges, the yellow ligand is the one without compound, the green one is the one without LTMP, and the red one is the standard version with all the components.}
  \label{fig: ablation}
\end{figure}

As shown in Figure \ref{fig: ablation}, without the intra-ligand long-range connection, the conformations were more likely to be unstable and have high energy. From the results in Table\ref{tab: ablation}, without the ligand-target compound, the conformations may not have reasonable poses, including the center position and orientation inside the pocket. With the LTMP feature assembler block, the ligand could better capture the 'shape' of the pocket by transferring the chemical and geometric messages of the target nodes to the ligand nodes. \\
%More visualizations are available in Appendix~\ref{app: visual}. \\

\begin{figure}[!ht]
  \centering
  \includegraphics[width=0.95\columnwidth]{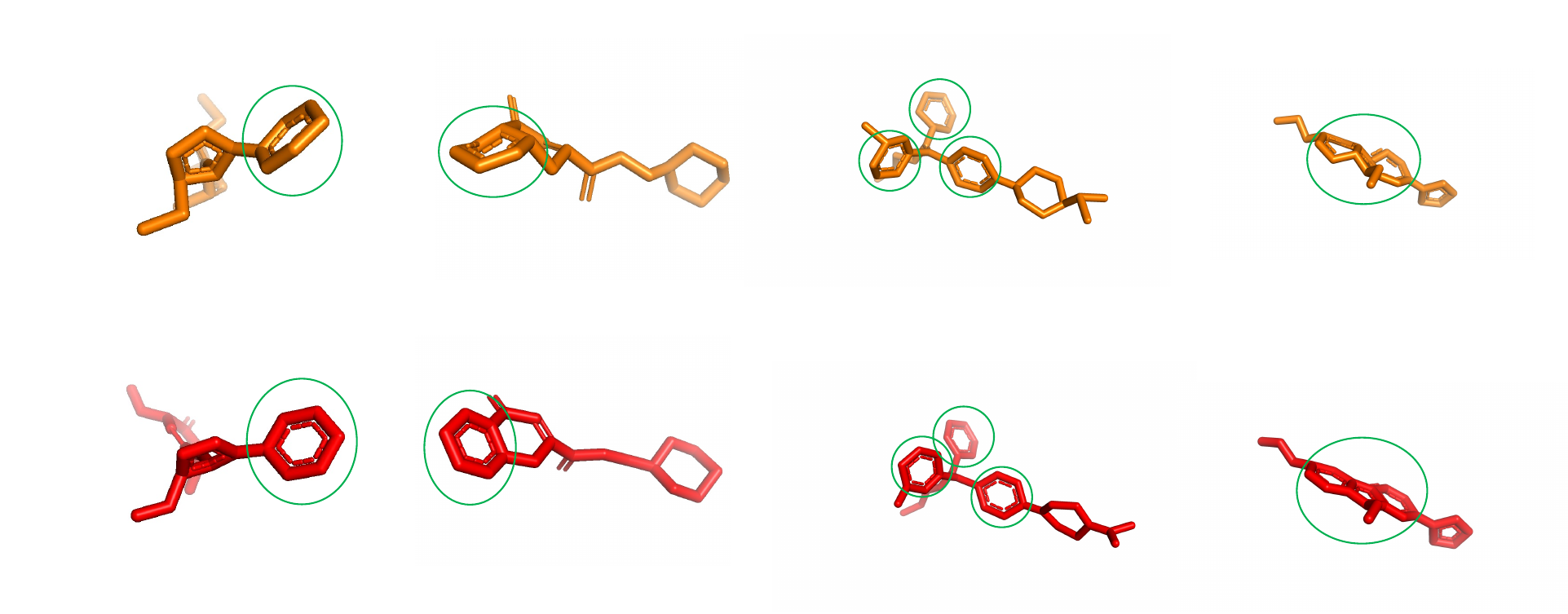}
  \caption{Ablation study for the effect of the guidance part. From left to right, the ligands are 5zjy, 5zk5, 6a6k, 6ggb. The red ligands are the ones with ligand property guidance while the orange ones are without guidance. The green circles point to the benzene rings in each ligand. Guidance helps to keep some geometric and chemical properties, such as the coplanarity of benzene rings.}
  \label{fig: ablation_guidance}
\end{figure}

Although the improvement in aligned RMSD and RMSD in Table~\ref{tab: ablation} for the guidance part was not significant, we further analyzed the results and found that guidance helps to maintain some geometric and chemical properties, such as the coplanarity of benzene rings, which may help to generate more 'reasonable' chemical molecules while satisfying energy or charge constraints. Such local structure constraints may not significantly change the overall structure, which is why the improvement in RMSD was not significant. We provide more details and analysis in Figure \ref{fig: ablation_guidance}.
\begin{table}
    \centering
    \begin{tabular}{c|cc|cc}
    \hline
    \multirow{2}*{Models}  & \multicolumn{2}{c|}{Aligned RMSD}   & \multicolumn{2}{c}{RMSD} \\
     \cline{2-5}
    ~ & mean & median  & mean & median \\
    \hline
    no compound construction &  2.72   &   1.63    &   2.17   & 2.97    \\
    no inter edges*  &  1.39E+33     &   23.12  & 1.39E+33 & 24.31    \\
    no LTMP  &  2.73  & 1.52 &  2.17 & 3.35 \\
    no guidance  & 2.65   & 1.418 & 2.05  & 3.11 \\
    \hline
    SIDEGEN & \textbf{2.629}  & \textbf{1.417}  & \textbf{2.033}  & \textbf{3.09}   \\
   
    \hline
    \end{tabular}
    \caption{Ablation study The training for no inter-edge version did not convergent finally and thus it fails.}
    \label{tab: ablation}
\end{table}

\subsection{Application on Drug-Target-Interaction Problem}
Our model can treat the DTI problem in an end-to-end manner without RDKit initialization.
To evaluate the binding pose for the generated conformations, we used the ligand RMSD in Eq.~\ref{eq: RMSD}. We also compared our method to recent docking tasks as baselines to assess the performance of our approach in generating biologically meaningful conformations that are consistent with the given conditions while also being relevant for drug design and development. The detailed results are in the \ref{app: ligand RMDF}. 

\section{Conclusion}
In this paper, we introduced SIDEGEN, a diffusion-based ligand conformation generation model that conditions on ligand molecular graph and target graph. SIDEGEN can generate conformations sampled from a random Gaussian distribution without any other ligand coordinate information. With the ligand-target compound construction, the generated conformations can be suitable for the target pockets, making them potentially useful in drug design projects. Moreover, the LTMP node feature assembler, which passes messages on a 2-node fully connected, directed, and self-loop graph containing nodes as ligand node features and the concatenated ligand-target node features, helps to capture the 'shape' of pockets.

Overall, SIDEGEN outperforms existing baseline methods and may be useful for drug design or conformation generation tasks. In the future, it may also be used in protein-protein docking projects or drug-protein soft docking projects.

\newpage
\onecolumn
\appendix
\section{Rot-Translation Invariant} \label{appen:se3 proof}
\begin{equation} \label{norm}
\tilde{\mathbf{X}}_L = \frac{\mathbf{X}_L - center_P}{\sqrt{var_P}}, 
\tilde{\mathbf{X}}_P = \frac{\mathbf{X}_P - center_P}{\sqrt{var_P}}
\end{equation}

\begin{equation} \label{norm_back}
\mathbf{X}_{L_0} = \tilde{\mathbf{X}}_{L_0} * \sqrt{var_P} + center_P
\end{equation}
\paragraph{Rot-translate Invariant}
In Definition~\ref{def:se3}, we consider two transformations: translation $T_g$ and rotation $R_g$. The translation invariance is guaranteed by transferring the ligand objects to zero mass at each time step. This ensures that the diffusion process operates on the center of mass of the ligand molecule, rather than its absolute position, enabling us to generate conformations that are translation invariant. Similarly, the rotation invariance is described in Theorem~\ref{thm:se3 markov chain}. \\

% By considering the SE(3) group of transformations, we can ensure that the diffusion process operates on a rotation-invariant representation of the ligand molecule. \\
\begin{definition}\label{def:se3}
A function f is equivariant to a set of transformations G, if for any $g \in G$, f and g commutes, i.e., $gf(x) = f(gx)$.
\end{definition}
\begin{theorem} \label{thm:se3 markov chain}
If the initial density $p(\mathbf{X}_{L_T})$ after normalization is rot-translate invariant, and the morkov kernel $p(\mathbf{X}_{L_{t-1}}|\mathbf{X}_{L_t}, \mathscr{G}_P, \mathscr{G}_L)$ is rotational invariant. Then the final density $p_\theta (\mathbf{X}_{L_0})$ is also rotational invariant.
\end{theorem}
The CoM-free transformation ensures that the initial density $p(\mathbf{X}_{L_T})$ is rot-translate invariant, as it is a standard Gaussian distribution \cite{com-free}. This is important for generating conformations that are consistent with the given conditions while also being rotation and translation invariant.

To guarantee the rot-translate equivariance of the Markov kernels $P(\mathbf{X}_{L_{t-1}}|\mathbf{X}_{L_t}, \mathscr{G}_P, \mathscr{G}_L)$, we use the edge feature and perform an equivariance transformation inspired by GeoDiff \cite{GeoDiff}. We transform all $\mathbf{X}_t$ with time steps t from 0 to T into CoM-free systems by subtracting the center of mass at each step, as shown in Theorem~\ref{thm:se3 kernel}. This ensures that the diffusion process operates on a consistent representation of the ligand molecule, enabling us to generate conformations that are both translation and rotation invariant.\\
\begin{theorem} \label{thm:se3 kernel}
The noise vector fields $\mathbf{s}_{\theta}(X, \mathscr{G}_P, \mathscr{G}_L. t)$ for the Markov kernels $P(\mathbf{X}_{L_{t-1}}|\mathbf{X}_{L_t}, \mathscr{G}_P, \mathscr{G}_L)$ are rotational equivariant. Formally, $R_g\mathbf{X}_L, \mathbf{F}_{L}= \Phi_{\theta}(R_g\mathbf{X}_L, R_g\mathbf{X}_{input}, \mathbf{F}_{L})$
\end{theorem}

For the guidance model based on EGNN \cite{EGNN}, we claim the following: 

\begin{theorem} \label{thm:guidance se3}
The energy model based on EGCL is rot-translate invariant with CoM cooperation. 
\end{theorem}

\begin{theorem} \label{thm:all se3}
If both the generation model $p(\mathbf{X}_{L_{t-1}}|\mathbf{X}_{L_t}, \mathscr{G}_P, \mathscr{G}_L)$ and the guidance model $G_{prop}(\mathscr{G}_L, c, t)$ are rotational invariant, then sampling from $p_{\theta} (\mathbf{X}_{L_0})$ is also rotational invariant. 
\end{theorem}

Together with Theorem.~\ref{thm:se3 markov chain}, Theorem.~\ref{thm:se3 kernel}, Theorem.~\ref{thm:all se3} and the CoM transmission, SIDEGEN is rot-translate invariant. \\

\subsection{Proof of Theorem.~\ref{thm:se3 markov chain}}

If the initial density $p(X_{L_T})$ after normalization is rot-translate invariant invariant, and the morkov kernel $p(X_{L_{t-1}}|X_{L_t}, \mathscr{G}_P, \mathscr{G}_L)$ is rotational invariant. Then the final density $p_\theta (X_{L_0})$ is also rotational invariant. \\

\begin{proof}

\begin{align}
    p_\theta (R_g(X_{L_0})) & = \int p(R_g(X_{L_T})p_\theta (R_g(X_{L_0:T-1})|R_g(X_{L_T})))d\mathbf{x_{1:T}} \\
    & = \int p(R_g(X_{L_T})\prod_{t=1}^{T}p_\theta (R_g(X_{L_{t-1}})|R_g(X_{L_t})))d\mathbf{x_{1:T}} \\
\end{align}
The initial density $p(X_{L_T})$ after normalization is rot-translate  invariant, gives $p(X_{L_T}) = p (R_g(X_{L_T}))$ \\
the morkov kernel is rotational invariant , gives $p(X_{L_{t-1}}|X_{L_t}) = (R_g(X_{L_{t-1}})|R_g(X_{L_t}))$, then \\
\begin{align}
    p_\theta (R_g(X_{L_0})) & = \int  p(X_{L_T})\prod_{t=1}^{T}p_\theta ((X_{L_{t-1}})|X_{L_t})d\mathbf{x_{1:T}} \\
     & = \int  p(X_{L_T})p_\theta ((X_{L_{0:T-1}})|X_{L_T})d\mathbf{x_{1:T}} \\
     & = p_\theta (X_{L_0})
\end{align}
\end{proof}

\subsection{Proof of Theorem.~\ref{thm:se3 kernel}}

The noise vector fields $\mathbf{s}_{\theta}(\mathbf{X}, \mathscr{G}_P, \mathscr{G}_L. t)$ for the Markov kernels $P(\mathbf{X}_{L_{t-1}}|\mathbf{X}_{L_t}, \mathscr{G}_P, \mathscr{G}_L)$ are rotational equivariant. \\
Formally, 
\begin{equation} \label{eq: se3 kernel}
R_g\mathbf{X}_L, \mathbf{F}_{L}= \Phi_{\theta}(R_g\mathbf{X}_L, R_g\mathbf{X}_{input}, \mathbf{F}_{L})
\end{equation}

\begin{proof}
In the ligand feature extractor, $\mathbf{F}_{L_{j}}$ and $\mathbf{E}_{jj'}$ are already invariant, the distance $D_{jj'}$ is a scalar, which is also invariant, \\
so for Eq.~ \ref{eq: feat_extractor1} \ref{eq: feat_extractor2}  \ref{eq: complex_gnnm} \ref{eq: complex_gnnh}:
\begin{equation} 
    R_g\mathbf{m}_{jj'} = \Phi_m(R_g\mathbf{F}_{L_{j}}^l, R_g\mathbf{F}_{L_{j'}}^l, R_g\mathbf{D}_{jj'}, R_g\mathbf{E}_{jj'};\theta_m) = \Phi_m(\mathbf{F}_{L_{j}}^l, \mathbf{F}_{L_{j'}}^l, \mathbf{D}_{jj'}, \mathbf{E}_{jj'};\theta_m) = \mathbf{m}_{jj'}
\end{equation}
and 
\begin{equation}
    R_g\mathbf{F}_{L_{j}}^{l+1} = \Phi_h(R_g\mathbf{F}_{L_{j}}^l, \sum_{j' \in N(j)}R_g\mathbf{m}_{jj'};\theta_h) = \Phi_h(\mathbf{F}_{L_{j}}^l, \sum_{j' \in N(j)}\mathbf{m}_{jj'};\theta_h) = \mathbf{F}_{L_{j}}^{l+1}
\end{equation}
In the target feature extractor, $f_{chem_i}^{l}, f_{geom_ii'}^{l}$ are scalars, and also invariant, for for Eq.~ \ref{eq: feat_extractor3}:
\begin{equation}
    R_g\mathbf{F}_{P_{i}}^{l+1} = \Phi_p(R_gf_{chem_i}^{l}, R_gf_{geom_ii'}^{l}) = \Phi_p(f_{chem_i}^{l}, f_{geom_ii'}^{l}) = \mathbf{F}_{P_{i}}^{l+1}
\end{equation}
The feature assembler block only updates the node features, which are invariant, so for Eq.~\ref{eq: LTMP}, \ref{eq: eq-out}
\begin{equation}
   R_g\mathbf{F}_{C}^{l+1} = LTMP(R_g\mathbf{F}_L, R_g\mathbf{Z}) = LTMP(\mathbf{F}_L, \mathbf{Z}) = \mathbf{F}_{C}^{l+1}
\end{equation}
where $\mathbf{Z} = Concat(\mathbf{F}_L, \mathbf{F}_P)$, $\mathbf{E}$ is the edge features for the ligand-target compound. Similar to the compound branch in Eq.~\ref{eq: complex_out}.
\begin{equation}
   R_g\mathbf{F}_{L_{out}} = Concat(AdaptiveAveragePool(R_g\mathbf{F}_{C}), R_g\mathbf{E}) = Concat(AdaptiveAveragePool(\mathbf{F}_{C}), \mathbf{E}) = \mathbf{F}_{L_{out}}
\end{equation}
Finally, for the edge-to-node equivariant transformation in Eq.~\ref{eq: eq-trans}
\begin{equation}
    R_g\mathbf{X}_{L_j}^{l+1} = \sum_{j' \in N(j)}R_g\frac{1}{D_{jj'}}(R_g\mathbf{X}_{input_j} - R_g\mathbf{X}_{input_j'})R_g\mathbf{F}_{L_{out_{jj'}}} =  R_g\sum_{j' \in N(j)}dir_{jj'}\mathbf{F}_{L_{out_{jj'}}} = R_g\mathbf{X}_{L_j}^{l+1}
\end{equation}
Therefore Eq.~ \ref{eq: se3 kernel} is satisfied.
\end{proof}

\subsection{Proof of Theorem.~\ref{thm:guidance se3}}

The energy model based on EGCL is rot-translate invariant with CoM 
\begin{proof}
As the EGCL formulas shown in Eq.~\ref{eq:egnn}, the transition equivariance is satisfied by applying CoM. We show the rotation equivariance here. \\
With rotation $R_g$, we will prove the model satisfies $$R_g\mathbf{X}_{L_j}, \mathbf{F}_{L_j} = EGCL(R_g\mathbf{X}_{L_j}, \mathbf{F}_{L_j})$$
\begin{equation} 
   m_{jj'}  = \Phi_m(\mathbf{F}_{L_j}^l,\mathbf{F}_{L_j'}^l,\mathbf{D}_{jj'}^2, \mathbf{E}_{jj'}) \\
   = \Phi_m(\mathbf{F}_{L_j}^l,\mathbf{F}_{L_j'}^l,\|R_g\mathbf{X}_{L_j}^{l} - R_g\mathbf{X}_{L_j'}^{l}\|^2, \mathbf{E}_{jj'}) \\
\end{equation}
Where $\|R_g\mathbf{X}_{L_j}^{l} - R_g\mathbf{X}_{L_j'}^{l}\|^2 = (\mathbf{X}_{L_j}^{l} - \mathbf{X}_{L_j'}^{l})^TR_g^TR_g(\mathbf{X}_{L_j}^{l} - \mathbf{X}_{L_j'}^{l}) = \|\mathbf{X}_{L_j}^{l} - \mathbf{X}_{L_j'}^{l}\|^2 = \mathbf{D}_{jj'}^2$, so 
\begin{equation} 
   m_{jj'} = \Phi_m(\mathbf{F}_{L_j}^l,\mathbf{F}_{L_j'}^l,\mathbf{D}_{jj'}^2, \mathbf{E}_{jj'}) \\
\end{equation}
Then, 
\begin{equation} 
    \mathbf{F}_{L_j}^{l+1}=\Phi_{\mathbf{h}}(\mathbf{F}_{L_j}^l, \sum_{j\neq j'}w_{jj'}m_{jj'}), \\
\end{equation}

\begin{equation} 
   R_g\mathbf{X}_{L_j}^{l+1} = R_g\mathbf{X}_{L_j}^{l} + \sum_{j\neq j'}R_g\frac{\mathbf{X}_{L_j}^{l} - \mathbf{X}_{L_j'}^{l}}{\sqrt{\mathbf{D}_{jj'}^2} + 1}
    \Phi{x}(\mathbf{F}_{L_j}^l,\mathbf{F}_{L_j'}^l,
    \mathbf{D}_{jj'}^2,\mathbf{E}_{jj'})
\end{equation}
Then, the energy model is rotational invariant.
\end{proof}

\subsection{Proof of Theorem.~\ref{thm:all se3}}

If both the generation model $p(\mathbf{X}_{L_{t-1}}|\mathbf{X}_{L_t}, \mathscr{G}_P, \mathscr{G}_L)$ and the guidance model $G_{prop}(\mathscr{G}_L, c, t)$ are rotational invariant, then sampling from $p_{\theta} (\mathbf{X}_{L_0})$ is also rotational invariant. 

\begin{proof}
$G_{prop}(\mathscr{G}_L, c, t)$ is rotational invariant gives that $$G_{prop}(R_g\mathbf{X}_L, \mathbf{F}_{L_j}^l, c, t) = G_{prop}(\mathbf{X}_L, \mathbf{F}_{L_j}^l, c, t)$$
Take derivatives and multiply $R_g$ on both sides, 
$$\nabla_{\mathbf{X}_L}G_{prop}(R_g\mathbf{X}_L, \mathbf{F}_{L_j}^l, c, t) = R_g\nabla_{R_g\mathbf{X}_L}G_{prop}(\mathbf{X}_L, \mathbf{F}_{L_j}^l, c, t)$$
\begin{align} 
   d\mathbf{X}_L & = [f(N_L, c, t)\mathbf{X}_Ldt + g(t)^2(\mathbf{s}_t(\mathbf{X}_L, N_L, c, t) \nonumber\\
   & + \lambda_{energy}\nabla_{\mathbf{X}_L}G_{energy}(\mathscr{G}_L, c, t) \nonumber\\
   & + \lambda_{gap}\nabla_{\mathbf{X}_L}G_{gap}(\mathscr{G}_L, c, t) \nonumber\\
   & + \lambda_{charge}\nabla_{\mathbf{X}_L}G_{charge}(\mathscr{G}_L, c, t))dt] \nonumber\\
   & + g(t)\overline{\mathbf{\omega_{\mathbf{X}_L}}} 
\end{align}
Then together with Theorem.~\ref{thm:se3 markov chain}, $p_{\theta} (\mathbf{X}_{L_0})$ is also rotational invariant.
\end{proof}

\section{Model Details}

\subsection{Hyperparameters} \label{appendix:Hyperparameters}
The essential hyperparameters are shown in Table.~\ref{Hyperparameters}.
\begin{table}[!htbp]
    \caption{Search space for SIDEGEN to perform well on the validation set. The best choices for hyperparameters are marked in \textbf{bold}.}
    \label{Hyperparameters}
    \centering
   % \resizebox{0.9\columnwidth}{!}{
    \begin{tabular}{cc}
    \hline
        PARAMETERS & SEARCH SPACE \\
        \hline
        Atom Type Num (Protein) & 6, 28, \textbf{32}\\
        Atom Type Num (Ligand) &  \textbf{28}\\
        Inter-edge Distance Cutoff & 2, 2.8, 5, 7, \textbf{8}, 10, 15\\
        Intra-edge Distance Cutoff & 2, \textbf{2.8}, 5, 7, 8, 10, 15\\
        Protein Downsampling Rate & 0.01, \textbf{0.03}, 0.05, 0.1, 1\\
        LTMP Depth & 1, \textbf{2}, 4, 6, 8 \\
        Training compound loss rate & \textbf{1}, 0.8, 0.5, 0.4, 0.1, \textbf{0} \\
        Learning Rate & \textbf{1e-3}, 1e-4, 1e-5 \\
        Learning Rate Scheduler & Cosine annealing \\
        Time steps & 1000, 5000 \\
        \hline
    \end{tabular}
\end{table}    

\subsection{LTMP} \label{appendix:LTMP}
\begin{figure}
     \centering
      \begin{subfigure}[b]{0.15\textwidth}
         \centering
         \includegraphics[width=\textwidth]{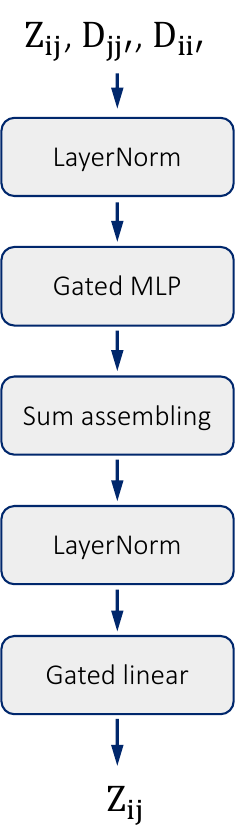}
         \caption{D to Z}
         \label{fig:D2z}
     \end{subfigure}
     \begin{subfigure}[b]{0.15\textwidth}
         \centering
         \includegraphics[width=\textwidth]{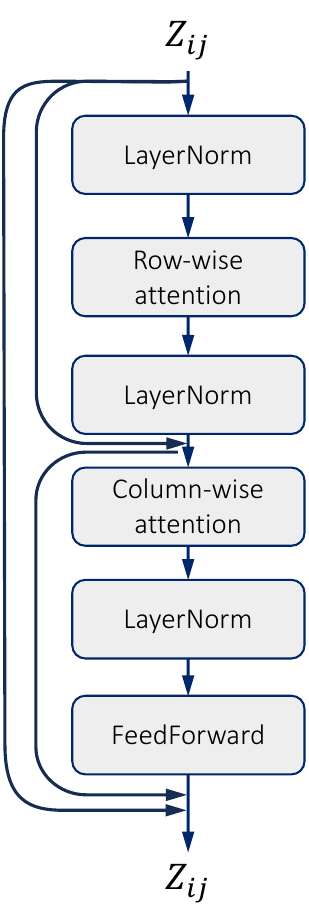}
         \caption{Z to Z}
         \label{fig:z2z}
     \end{subfigure}
     \begin{subfigure}[b]{0.15\textwidth}
         \centering
         \includegraphics[width=\textwidth]{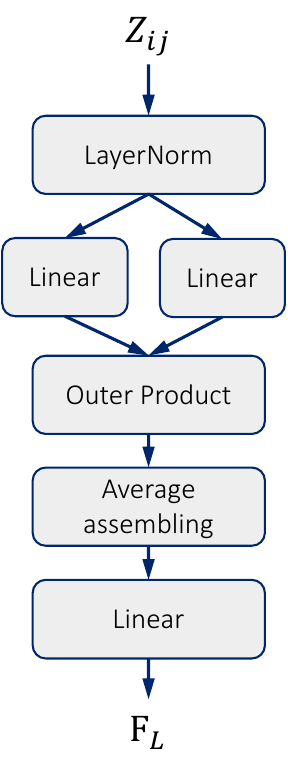}
         \caption{Z to L}
         \label{fig:z2l}
     \end{subfigure}
     \begin{subfigure}[b]{0.15\textwidth}
         \centering
         \includegraphics[width=\textwidth]{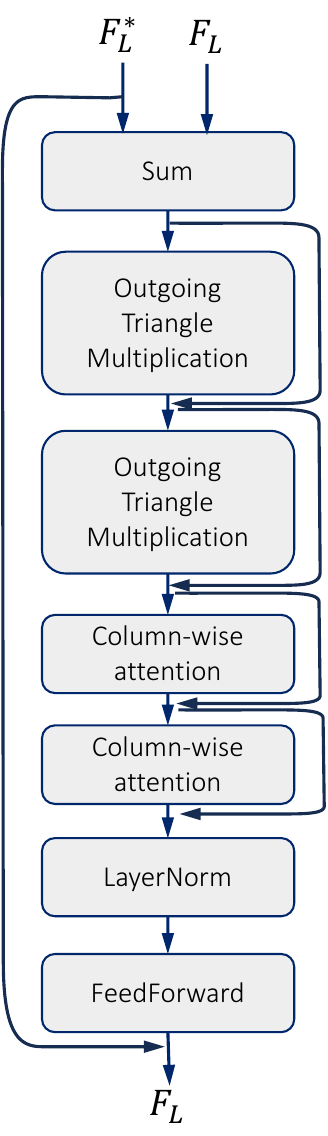}
         \caption{L to L}
         \label{fig:l2l}
     \end{subfigure}
     \begin{subfigure}[b]{0.15\textwidth}
         \centering
         \includegraphics[width=\textwidth]{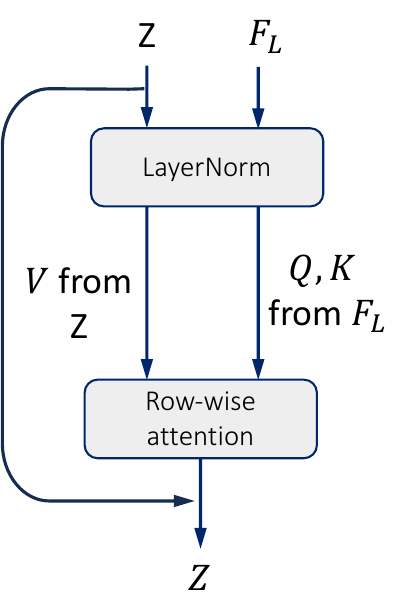}
         \caption{L to Z}
         \label{fig:l2Z}
     \end{subfigure}
     \begin{subfigure}[b]{0.15\textwidth}
         \centering
         \includegraphics[width=\textwidth]{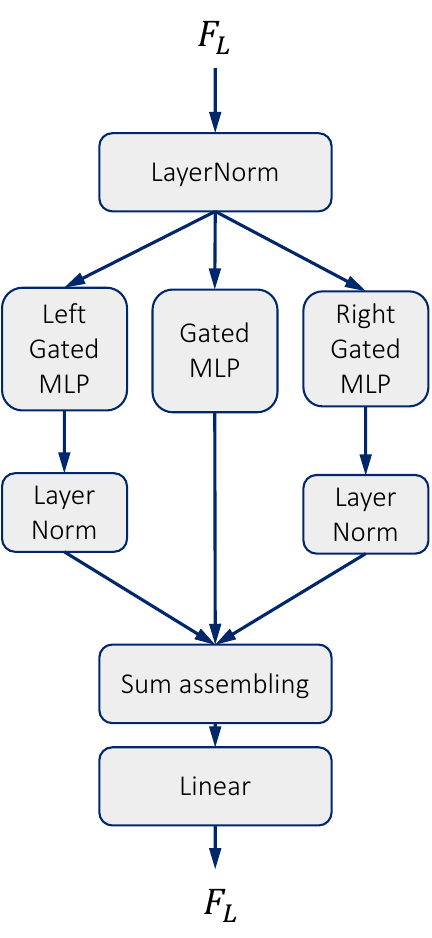}
         \caption{Trigonometric Multiplication}
         \label{fig:tri_mul}
     \end{subfigure}
    \caption{Sub-blocks of LTMP: Z to L, Z to Z, L to Z, D to Z, and L to L. The last subgraph shows the trigonometric multiplication block in the L to L sub-block.}
    \label{fig:evo_subblocks}
\end{figure}

The LTMP feature assembler considers the ligand and compound graph as two nodes of a directed self-looped graph and tries to pass massages inside the graph. It consists of 5 sub-blocks as shown in Figure \ref{fig:encoder}(b): D to Z, Z to Z, Z to L, L to L, and L to Z. The detailed structures of these 5 blocks are shown in Figure \ref{fig:evo_subblocks}. \\

\subsection{Graph representation} \label{appen:graph repr}
\paragraph{Ligand graph}
Ligand graphs have nodes as the heavy atoms with node features $F_L \in \mathbb{R}^{d_l\times n}$ and edges being the chemical bonds with edge features $\mathbf{E_{{jj'}_{local}}}$. The node features are one-hot embedded from 28 atom types while the edge features are embedded by edge types and Euclidean distances. \\
\paragraph{Target graph}
The node features for the target graph consist of two components: chemical features and geometric features. The chemical features include 32 node types and the trainable chemical properties for the neighboring K atoms (K=16). To better encode the 'shape' of the pocket surface, trainable geometric features including the Gaussian curvatures and the mean curvatures are also embedded in the node features.
\subsection{Feature extractor}
\begin{figure}
     \centering
     \begin{subfigure}[b]{0.45\textwidth}
         \centering
         \includegraphics[width=0.45\textwidth]{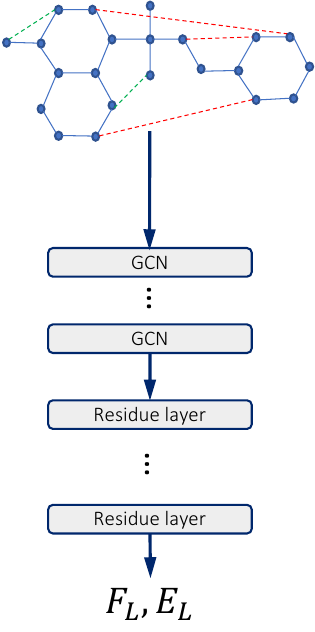}
         \caption{Ligand feature extractor}
         \label{fig:lig_feat}
     \end{subfigure}
     \begin{subfigure}[b]{0.45\textwidth}
         \centering
         \includegraphics[width=0.45\textwidth]{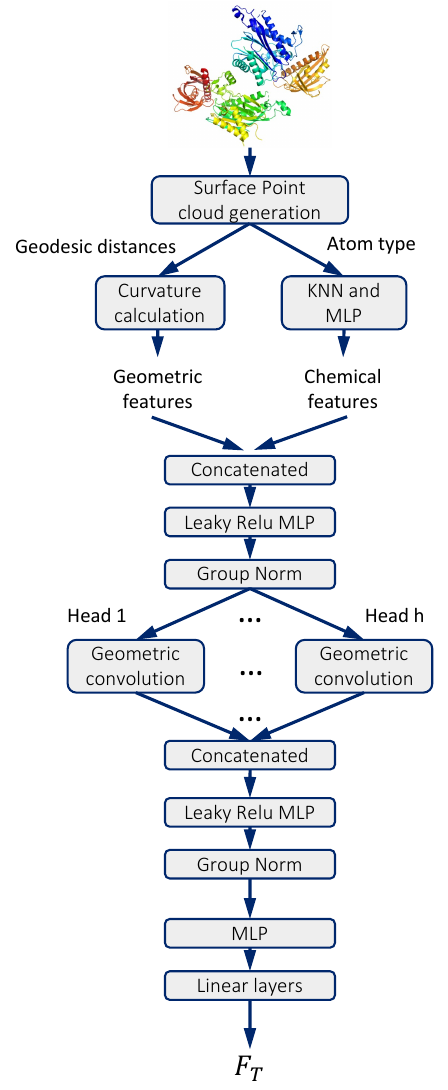}
         \caption{Target feature extractor}
         \label{fig:tar_feat}
     \end{subfigure}
\end{figure}

We try two combinations of backbone graph neural networks for the ligand feature extractor. The first one is Graph Convolution Network (GCN) for both ligand-target interaction and compound branches. The second one is SchNet \cite{schnet} for compound branch and Graph Isomorphism Network (GIN) for ligand-target interaction. The detailed structure is shown in Figure \ref{fig:lig_feat}. We also try a model similar to the energy model based on the EGNN model \cite{EGNN, EGNN_gene} with the ligand atom types fixed and without the output MLP layer. The results show that the GCN version is better, so we finally it.\\
For the target graph, we choose the differentiable geodesic convolution-based surface point cloud feature extractor Dmasif, the detailed structure is shown in Figure \ref{fig:tar_feat}.\\

\subsection{Energy Model} \label{appendix:guidance}
The energy model for guiding sampling is designed to predict the chemical properties as shown in Section \ref{sec:guided_diffusion}. The input of the energy model is the ligand molecular graph together with their chemical properties calculated by Psi4 \cite{Psi411,Psi414} and RDkit \cite{rdkit}. The first model we try is a similar structure to the ligand extractor (GCN based) except for the output MLP layer. However, the time spent is relatively long with training 2000 epochs taking 6 days on 1 Tesla A100 GPU. To save time, we also try the EGNN model \cite{EGNN, EGNN_gene} with the ligand atom types fixed. The performances between to two models are similar to the energy model but save more time. Therefore, the EGNN-based model is finally selected with the equivariant convolution layer as shown in \ref{eq:egnn}. \\
\paragraph{Guidance model Loss Function}
The guidance model is pre-trained for each of the chemical properties. For each of them, the loss function is designed as the L1-loss between the predictions and the labels as shown in Eq.~\ref{eq:guidance_loss}, where $G_{\theta'}$ denotes the parameterized prediction of the chemical properties by the guidance model and $c_{prop}$ denotes the ground truth chemical properties. \\

\begin{equation} \label{eq:guidance_loss}
    \mathcal{L} = \mathbb{E}|G_{\theta'} - c_{prop}|
\end{equation}

\begin{equation} 
\begin{split}\label{eq:egnn}
   & m_{jj'} = \Phi_m(\mathbf{F}_{L_j}^l,\mathbf{F}_{L_j'}^l,\mathbf{D}_{jj'}^2, \mathbf{E}_{jj'}), 
    w_{jj'}=\Phi_w{m_{jj'}}, 
    \mathbf{F}_{L_j}^{l+1}=\Phi_{\mathbf{h}}(\mathbf{F}_{L_j}^{l}, \sum_{j\neq j'}w_{jj'}m_{jj'}), \\
   & \mathbf{X}_{L_j}^{l+1} = \mathbf{X}_{L_j}^{l} + \sum_{j\neq j'}\frac{\mathbf{X}_{L_j}^{l} - \mathbf{X}_{L_j'}^{l}}{\sqrt{\mathbf{D}_{jj'}^2} + 1}
    \Phi{x}(\mathbf{F}_{L_j}^l,\mathbf{F}_{L_j'}^l,
    \mathbf{D}_{jj'}^2,\mathbf{E}_{jj'})
\end{split}
\end{equation}

Here, $\Phi_w, \Phi_m, \Phi_x, \Phi_h$ are learnable networks, $m_{jj'}$ is the message, $\mathbf{F}_{L_j}^l$ is the ligand node feature consisting of node types, time, and chemical properties. $\mathbf{D}_{jj'}$ is the Eulidean distance and $\mathbf{E}_{jj'}$ is the edge feature, which is the chemical bond type. \\
Moreover, the transition equivariance is guaranteed by the CoM similar to the main model and the model is rotational invariant. \\

\paragraph{Experiments settings}
Three separate guidance models for gaps, energy, and charges were trained separately. Each model was trained on one Tesla A100 GPU for five days for 5000 epochs. The learning rate was set to be $2e-4$ with a weight decay of $1e-16$. We calculated the Self-consistent field (SCF) energy and molecular orbital (HOMO)–lowest unoccupied molecular orbital (LUMO) energy gaps using the Psi4 software \cite{Psi411} and the Marsili-Gasteiger Partial Charges using RDKit \cite{rdkit}.\\

\section{More results} \label{app: more-results}
GeoDiff is trained on  GEOM-QM9 \cite{qm9} and GEOM-Drugs \cite{drugs} datasets, without any protein data inside them. Our model requires target information thus the above datasets are not available. We test the model weights given by GeoDiff and also retrain it on the PDBBind-2020 dataset. The direct testing on the given weights does not convergent for most of the ligands in the PDBBind datasets. 

\section{Application on Drug-Target-Interaction Problem} \label{app: ligand RMDF}
As shown in \ref{tab:rmsd}, without any extra optimization, our model achieves comparable results compared to the traditional method (GNINA \cite{gnina} and GLIDE (c.) \cite{glide} and the deep learning method (EquiBind \cite{equibind} and TankBind \cite{tankbind}). With a simple one-step empirical force field (FF) \cite{ff} optimization, our method outperforms most of the existing methods or their combination of median and 75th quantile.\\
% \paragraph{RMSD}
\begin{table}
    \centering
    \begin{tabular}{c|ccc}
    \hline
    \multirow{2}*{Models}  & \multicolumn{3}{c}{Ligand RMSD Percentiles($\mathring{\textrm{A}}$)$\downarrow$} \\
     \cline{2-4}
    ~ & 25th & 50th & 75th\\
    \hline
    GNINA & 2.4 & 7.7 & 17.9 \\
    GLIDE (c.)  & 2.6 & 9.3 & 28.1 \\
    \hline
    EquiBind & 3.8 & 6.2 & 10.3      \\
    TANKBind  & 2.4 & 4.28  & 7.5       \\
    \hline
    P2RANK+GNINA & \textbf{1.7} & 5.5 & 15.9 \\
    EQUIBIND+GNINA & 1.8 & 4.9 & 13 \\
    \hline
    *GeoDiff-PDBBind & 29.21 & 40.33 & 79.62 \\
    \hline
    SIDEGEN & 5.49 & 7.29 & 9.50        \\
    SIDEGEN + FF & 1.8 & \textbf{2.49} & \textbf{3.40}  \\
    \hline
    \end{tabular}
    \caption{Ligand RMSD on PDBBind-2020(filtered), Geodiff does not consider the position of ligands during docking, and centered the results to the origin of the Cartesian coordinate system.}
    \label{tab:rmsd}
\end{table}

There are no ethical issues.

\section*{Acknowledgments}

%Bibliography
\bibliographystyle{unsrt}

\end{document}